\documentclass[aps,pra,superscriptaddress,eqsecnum,onecolumn,nofootinbib]{revtex4}
\usepackage{color,graphicx,ulem}%umoline}
\usepackage{amsmath,amssymb}
\usepackage{graphicx}
\usepackage{braket}
\usepackage{times,dsfont,amssymb,amsmath,amsthm,amsfonts,amsbsy,mathrsfs,bm,bbm,graphicx,graphics,epsfig,multirow,mathtools,color,bbold,empheq}

\definecolor{notas}{cmyk}{0,1,0.50,0}%{0.1,0.6,1,0.07}

\newcommand{\LON}{\hbox{LON}}

\newcommand{\ketbra}[1]{\ket{#1}\hspace{-0.3em}\bra{#1}}
\newcommand{\trace}{\textrm{Tr}}
\newcommand{\ourmatrix}{\bm{S}}

\newcommand\tvdBS{D\big(P_{\textrm{BS}|\bm{n}_A,\bm{U}}, P_{\textrm{BS}|\bm{n}_A,\bm{L}}\big)}
\newcommand\bcBS{F\big(P_{\textrm{BS}|\bm{n}_A,\bm{U}}, P_{\textrm{BS}|\bm{n}_A,\bm{L}}\big)}
\newcommand\tvd{D\big(P_{\textrm{Q}|\bm{U}}, P_{\textrm{Q}|\bm{L}}\big)}
\newcommand\bc{F\big(P_{\textrm{Q}|\bm{U}}, P_{\textrm{Q}|\bm{L}}\big)}
\newcommand\bcN{F\big(P_{\textrm{Q}|N,\bm{U}}, P_{\textrm{Q}|N,\bm{L}}\big)}
\newcommand\qf{F\big(\rho_{AB|\bm{U}},\rho_{AB|\bm{L}}\big)}
\newcommand\qfN{F\big(\rho_{AB|N,\bm{U}},\rho_{AB|N,\bm{L}}\big)}
\newcommand\qtd{D\big(\rho_{AB|\bm{U}},\rho_{AB|\bm{L}}\big)}
\newcommand\fe{F_e\big(\rho_{\text{th},B},\mathcal{E}^{-1}_{\bm{U}}\circ\mathcal{E}_{\bm L}\big)}
\newcommand\Falpha{F(\bm{\alpha},\bm{\alpha}^\dagger)}
\newcommand\Galpha{G(\bm{\alpha},\bm{\alpha}^\dagger)}

\newcommand\cvar{\upsilon}

\begin{document}
\title{In situ characterization of linear-optical networks in randomized boson sampling}

\author{Saleh Rahimi-Keshari}
\email{s.rahimik@gmail.com}
\affiliation{Center for Quantum Information and Control, University of New Mexico, Albuquerque, New Mexico 87131-0001, USA}
\affiliation{Institut f\"ur Theoretische Physik, Leibniz Universit\"at Hannover, Hannover, Germany}
\affiliation{School of Physics, Institute for Research in Fundamental Sciences (IPM), P.O. Box 19395-5531, Tehran, Iran}
\affiliation{ICFO - Institut de Ciencies Fotoniques, The Barcelona Institute of Science and Technology, 08860 Castelldefels (Barcelona), Spain}
\affiliation{Integrated Quantum Optics, Paderborn University, Warburger Strasse 100, 33098 Paderborn, Germany}
\affiliation{Department of Physics, University of Tehran, P.O. Box 14395-547, Tehran, Iran
}
\affiliation{School of Nano Science, Institute for Research in Fundamental Sciences (IPM), P.O. Box 19395-5531, Tehran, Iran}

\author{Sima Baghbanzadeh}
\email{sima.baghbanzadeh@gmail.com}
\affiliation{Center for Quantum Information and Control, University of New Mexico, Albuquerque, New Mexico 87131-0001, USA}
\affiliation{School of Physics, Institute for Research in Fundamental Sciences (IPM), P.O. Box 19395-5531, Tehran, Iran}
\affiliation{Integrated Quantum Optics, Paderborn University, Warburger Strasse 100, 33098 Paderborn, Germany}

\author{Carlton M. Caves}
\email{ccaves@unm.edu}
\affiliation{Center for Quantum Information and Control, University of New Mexico, Albuquerque, New Mexico 87131-0001, USA}

\begin{abstract}
We introduce a method for efficient, {\it in situ\/} characterization of linear-optical networks (\LON s) in randomized boson-sampling (RBS) experiments.  We formulate RBS as a distributed task between two parties, Alice and Bob, who share two-mode squeezed-\hbox{vacuum} states.  In this protocol, Alice performs local measurements on her modes, either photon counting or heterodyne.  Bob implements and applies to his modes the \LON\ requested by Alice; at the output of the \LON, Bob performs photon counting, the results of which he sends to Alice via classical channels.  In the ideal situation, when Alice does photon counting, she obtains from Bob samples from the probability distribution of the RBS problem, a task that is believed to be classically hard to simulate.  When Alice performs heterodyne measurements, she converts the experiment to a problem that is classically efficiently simulable, but more importantly, enables her to characterize a lossy \LON\ on the fly, without Bob's knowing and without changing anything at Bob's end (this is what we mean by {\it in situ}).  We introduce and calculate the fidelity between the joint states shared by Alice and Bob after the ideal and lossy {\LON}s as a measure of distance between the two {\LON}s.  Using this measure, we obtain an upper bound on the total variation distance between the ideal probability distribution for the RBS problem and the probability distribution achieved by a lossy \LON.  Our method displays the power of the entanglement of the two-mode squeezed-vacuum states: the entanglement allows Alice to choose for each run of the experiment between RBS and a simple characterization protocol based on first-order coherence
between complex amplitudes.
\end{abstract}

\maketitle

\section{Introduction}
\label{sec:intro}

It is strongly believed that quantum computers can perform certain computational tasks much faster than classical computers.  A universal fault-tolerant quantum computer, however, is still not available, so there is keen interest in intermediate models of quantum computation, which can demonstrate quantum-computational speedups~\cite{Preskill2012} with simpler physical systems and algorithms~\cite{Harrow2017}.  A class of these intermediate models consists of sampling problems, i.e., generating samples from the output probability distribution of a quantum circuit that is believed to be hard to simulate efficiently classically~\cite{AA,Bremner2016,Gao2017,Boxio2018}.

Just such a sampling problem is boson sampling, which has attracted much attention due to its simple physical implementations~\cite{AA}.  In boson sampling, single photons are injected into $N$ input ports of an $M$-mode (passive) linear-optical network (LON) that is described by an $M\times M$ Haar-random unitary transfer matrix, and one samples from the output photon-counting probability distribution. This sampling task is strongly believed to be classically hard to simulate, and hence it is proposed as a candidate to demonstrate quantum-computational speedups in the near future. This has also led to small-scale experimental demonstrations of boson sampling~\cite{Broome2013,SpringR2013,Spagnolo2014,Tillmann2013,Carolan2015, Wang2017,Loredo2017,He2017}.

Randomized boson sampling (RBS)~\cite{RBS} is a generalized version of boson sampling in which single photons are injected into random input ports of a Haar-random \LON.  RBS is generally believed to be classically hard to simulate.  The standard protocol for RBS uses two-mode squeezed-vacuum states, which are less demanding to prepare than the single-photon states required for boson sampling.  RBS has also been experimentally investigated in small-scale experiments~\cite{Bentivegna2015,Zhong2018}.

One important question confronting any sampling is that of how to verify efficiently that the generated samples are indeed from the correct output probability distribution. A {\it verification test\/} is a classical algorithm that receives samples from a sampling experiment and returns {\bf yes} if they are drawn from a distribution sufficiently close to the correct probability distribution, which cannot be sampled efficiently classically, and {\bf no} otherwise~\cite{Aaronson2016}.  A verification test must not pass samples that are generated by an efficient classical algorithm. It is unlikely that an efficient verification test exists for classically hard sampling problems without additional promises and assumptions~\cite{Hangleiter2018}; even then, it is not clear what promises and assumptions are required.

Other approaches with a different objective~\cite{Aharonov2017,Broadbent2009, Aolita2015,Miller2017,Hangleiter2017,Broadbent2018,Fitzsimons2018,Takeuchi2018,Ferracin2018,Hayashi2019,Ferracin2018u} seek to {\it certify\/} the workings of a quantum computing device instead of trying to {\it verify\/} answers to a problem (here samples from a probability distribution). These approaches are usually formalized in terms of a two-party scenario, a certifier with limited quantum capability and a prover, who supposedly has the quantum power, but can cheat.  The ultimate goal in certification approaches, which are based on some physical and/or computational hardness assumptions, is to certify the operation done by the prover and check whether the prover has been honest.

If, on the other hand, one assumes honesty between experimenters and assumes certain types of error (these are typical assumptions in physics experiments), another, more basic task in a quantum-supremacy experiment, more akin to process tomography, is to {\it characterize\/} fully the quantum device.  Preferably, characterization should be done {\it in situ}, i.e., while the experiment is running and without making any changes to the quantum device to perform the characterization. Characterization is the essential physical prerequisite for the more information-theoretic tasks of certification and verification.
Once one has the full characterization of the generally flawed actual device, one wants to have available performance measures that use the characterization to assess the validity of the experiment.

In this paper, we introduce an efficient method for {\it in situ\/} characterization of linear-optical networks (LON) in RBS experiments. Our characterization method is based on an interesting application of entanglement to convert RBS to a problem that enables efficient characterization of the \LON.  We formulate RBS as a distributed task between two parties, Alice and Bob, who share two-mode squeezed-vacuum states that are weakly squeezed.  To perform RBS, Alice counts photons on her half of the two-mode squeezed-vacuum states, Bob inputs the corresponding modes at his end into a Haar-random \LON\ that he has constructed based on instructions from Alice, and at the output of the \LON, Bob performs photon counting, the results of which he reports to Alice.  Because of the photon correlations in two-mode squeezed vacuum, when Alice counts a single photon (no photon) in her half of the state, a single photon (no photon) is injected into the corresponding port of Bob's~\LON.  The result is to input single photons into random, but heralded ports of the \LON; the photocounts that Bob reports to Alice are drawn from the output photocount distribution for whatever heralded input applies in each RBS~run.

If Bob has constructed the ideal \LON, Alice and Bob are sampling from the joint probability distribution of the RBS problem.  Generally, however, Bob's \LON~has losses; assuming that losses are the only kind of imperfection in the \LON, Alice can efficiently characterize the \LON\ by performing heterodyne measurements at her end, instead of photocounting.  The heterodyne measurements prepare the inputs to the \LON\ in coherent states.  As we show, Alice can use the first-order coherence (second moments) of the heterodyne outcomes, conditioned on the counts she receives from Bob, to characterize the \LON\ on the fly.  An important point is that the characterization runs are interspersed with the RBS runs, without Bob's knowing which is which and without making any changes to the apparatus at Bob's end.  In our protocol, we assume that Bob is honest in implementing the \LON\ as best he can, but at his end, there can be losses at the input of the \LON, within the \LON, and in the detectors, all of which can be modeled and absorbed into a lossy \LON.  Bob's inability to know which are the characterization runs is a way of saying that no changes are made to the \LON\ during the characterization runs.

Our characterization protocol can be thought of as a many-mode version of SU(1,1) interferometry~\cite{Yurke1986a}.  The connections of our protocol to SU(1,1) interferometry and, more generally, to quantum metrology are explored in~\cite{Caves2019a}.

We go on to introduce a measure of distance between {\LON}s, which is interesting in its own right and can be used as a performance measure for other quantum experiments with a \LON.  This measure is the fidelity between the joint state shared by Alice and Bob after the ideal \LON\ and the joint state after the actual, lossy \LON\ implemented by Bob.  By using this measure, Alice can obtain an upper bound on the total variation distance between the ideal probability distribution for the RBS problem and the probability distribution achieved in the experiment.  Thus, our {\it in situ\/} characterization procedure can check that a RBS device samples from a probability distribution that is close to the ideal photocount distribution.  We emphasize that this does not address the problem of verification of RBS, i.e., distinguishing the samples generated by the experiment from ones generated by some classical algorithm.

This paper focuses on {\it in situ\/} device characterization within the context of RBS; in future papers, we plan to address certification of RBS experiments and verification of boson-sampling experiments.  Section~\ref{sec:distributedRBS} describes the ideal RBS protocol cast as a distributed task between Alice and Bob.  Section~\ref{sec:characterization}, the heart of the paper, describes and analyzes the protocol for characterizing the \LON, first generalizing the description of RBS runs to lossy {\LON}s (Sec.~\ref{sec:samplingruns}), then showing how the statistics of the characterization runs can be used to determine the transfer matrix of a lossy \LON\ (Sec.~\ref{sec:characterizationruns}), and finally introducing a measure of distance between the ideal and lossy {\LON}s and showing how this performance measure can be used to bound the total variation distance between the ideal photocount distribution and the distribution achieved with a lossy \LON\ (Sec.~\ref{sec:comparing}).  Implications and generalizations of our work are explored in a concluding section (Sec.~\ref{sec:discussion}), which unlike most concluding sections, is worth reading for the several (at least three) important observations it makes.

Two appendices provide details that, though technical, are both important and instructive.  Appendix~\ref{sec:fidelitylowerbound} develops the formal description of lossy \LON s and gives the details of the bounds underlying the distance measures between ideal and lossy \LON s.  Appendix~\ref{sec:hetstats}
develops the full description of Alice's conditional heterodyne statistics and then considers what kind of \LON\ characterization can be achieved in situations where only intensity correlations are available, instead of the first-order (complex-amplitude) correlations used in our protocol.

%%%%%%%%%%%%%%%%%%%%%%%%%%%%%%%%%%%%%%%%%%%%%%%%%%%%%%%%%%%%%%%%%%%%%%%%%%
%%%%%%%%%%%%%%%%%%%%%%%%%%%%%%%%%%%%%%%%%%%%%%%%%%%%%%%%%%%%%%%%%%%%%%%%%%
%%%%%%%%%%%%%%%%%%%%%%%%%%%%%%%%%%%%%%%%%%%%%%%%%%%%%%%%%%%%%%%%%%%%%%%%%%
\section{Distributed RBS between two parties}
\label{sec:distributedRBS}

\begin{figure}
\centering
\includegraphics[width=0.7\columnwidth]{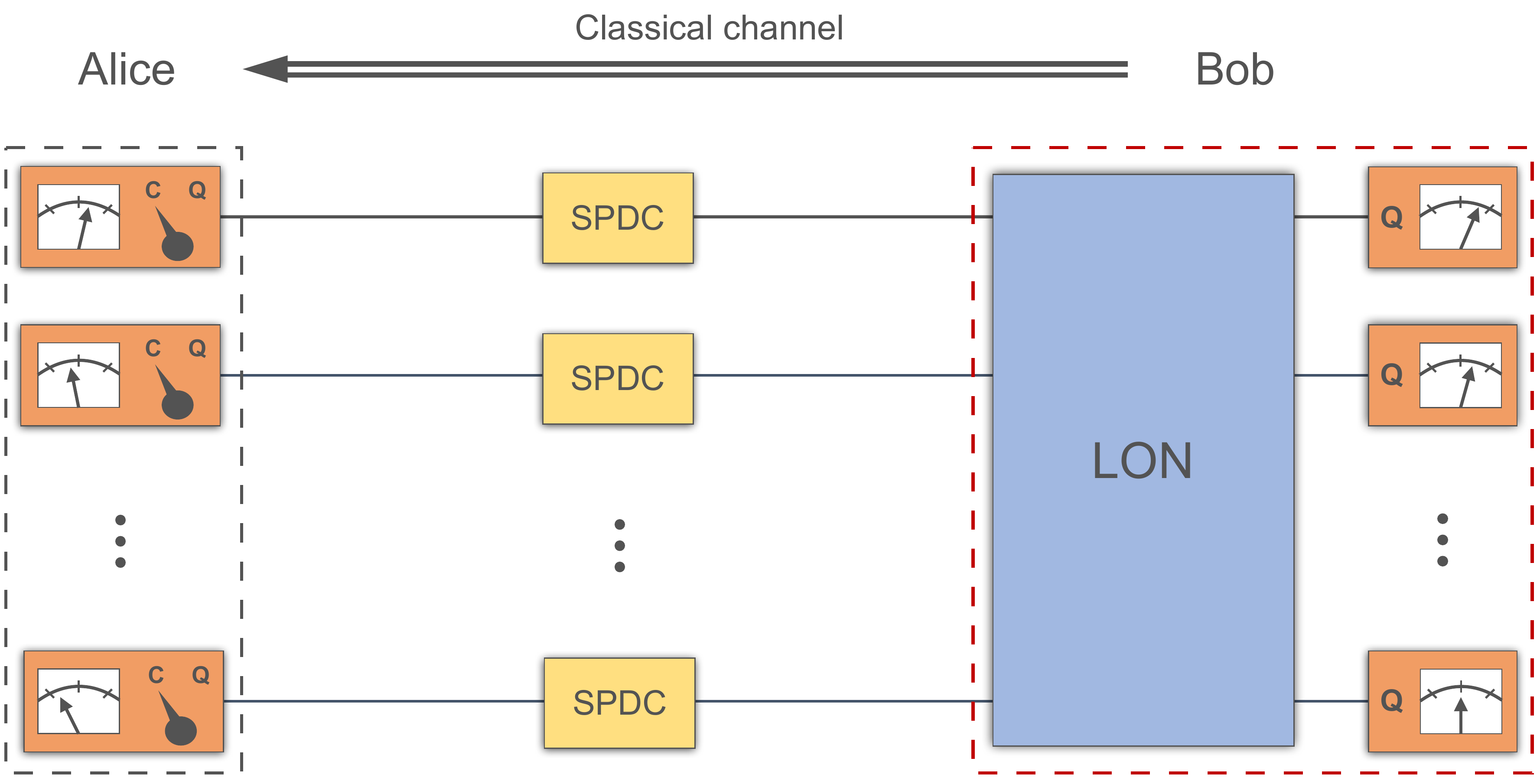}
\caption{$M$ two-mode squeezed-vacuum states with the same squeezing parameter are generated by SPDC sources and shared between Alice and Bob.  Bob inputs his modes to the \LON, makes photon-counting measurements at the output of the \LON, and sends the outcomes to Alice via a classical channel.  Alice uses photon-counting measurements, denoted by Q, for the sampling runs of the RBS problem and makes heterodyne measurements, denoted by C, for characterizing Bob's \LON.  Because of the entanglement shared between the two parties, Alice can switch from a problem, based on photon counting, that is classically hard and not useful for characterization to a problem, based on heterodyne measurements, that is classically efficiently simulable and can be used for characterization.}
\label{fig:RBSsetup}
\end{figure}

In the version of RBS we consider here (Fig.~\ref{fig:RBSsetup}), Alice and Bob share $M$ pairs of modes, with annihilation and creation operators $a_j$ and $a_j^\dagger$ at Alice's end and $b_j$ and $b_j^\dagger$ at Bob's end, $j=1,\ldots, M$.  Spontaneous parametric down-conversion (SPDC) sources generate identical two-mode squeezed-vacuum states in each pair of modes,
\begin{align}
\label{eq:2MSV}
	\ket{\psi_{AB}}=\sqrt{1-|\chi|^2} \sum_{n=0}^\infty \chi^n \ket{n}_A\otimes\ket{n}_B\,.
\end{align}
Here $\ket{n}_A$ and $\ket{n}_B$ are Fock states for the $A$ and $B$ modes output by a SPDC source, and $\chi$ is a complex number satisfying $|\chi|\le1$, which determines the amount and phase angle of the squeezing. It is convenient in what follows to set the relative phase of the $A$ and $B$ modes by
choosing $\chi$ to be real. The state for all $M$ modes is
\begin{align}\label{eq:PsiAB}
\ket{\Psi_{AB}}=\bigotimes_{i=1}^{M}\ket{\psi_{AB}}_i
=(1-\chi^2)^{M/2}\sum_{N=0}^\infty\chi^N
\hspace{-5pt}\sum_{{\scriptstyle{\bm{n}_A}\atop\scriptstyle{|\bm{n}_A|=N}}}\ket{\bm{n}_A}\otimes\ket{\bm{n}_B=\bm{n}_A}\,.
\end{align}
Here $\bm{n}_A=n_{A,1},\ldots,n_{A,M}$ and $\bm{n}_B=n_{B,1},\ldots,n_{B,M}$ are lists of the numbers of photons in the $A$ modes and $B$ modes; because of the entanglement in the two-mode squeezed vacuum, $\bm{n}_B=\bm{n}_A$, as indicated.  The total number of photons in the $A$ modes is denoted by
\begin{align}
|\bm{n}_A|=\sum_{i=1}^M n_{A,i},
\end{align}
and this is also the total number of photons in the $B$ modes, i.e., $|\bm{n}_A|=|\bm{n}_B|$.

Alice's modes ($A$ modes) are sent directly to photodetectors, and Bob's modes ($B$ modes) are directed to a \LON, described by a unitary operator $\mathcal{U}$, followed by photodetections.  The \LON\ is characterized completely by an $M\times M$ unitary {\it transfer matrix\/} $\bm U$ that describes the transformation of modal creation operators from input to output:
\begin{align}\label{eq:Ubtransform}
\mathcal{U}\,b_j^{\dagger}\;\mathcal{U}^\dagger=\sum_{k=1}^{M} U_{jk}b_k^{\dagger}\,.
\end{align}
By using this relation, it is straightforward to see that coherent states transform according to the transfer matrix,
\begin{align}\label{eq:Ubeta}
\mathcal{U}\ket{\bm{\beta}}=\ket{\bm{\beta}\bm{U}}\,.
\end{align}
where $\bm{\beta}=\begin{pmatrix}\beta_1&\beta_2&\cdots&\beta_M\end{pmatrix}$ is the row vector of coherent-state amplitudes.

In this setup, Alice and Bob sample from the joint probability distribution
\begin{align}
\label{eq:JointPrRBS}
P_{\textrm{Q}}(\bm{n}_A, \bm{n}_B|\bm{U})
=\big\vert \bra{\bm{n}_A,\bm{n}_B}(\mathcal{I}_A\otimes\mathcal{U}) \ket{\Psi_{AB}} \big\vert^2
=(1-\chi^2)^M \chi^{2|\bm{n}_A|}\,\big\vert \bra{\bm{n}_B}\mathcal{U} \ket{\bm{n}_A} \big\vert^2\,,
\end{align}
where $\bm{n}_A$ and $\bm{n}_B$ here denote counts in the $A$ and $B$ modes.  Because $\mathcal{U}$ is photon-number preserving, the probability is zero unless $|\bm{n}_B|=|\bm{n}_A|$.  Here and throughout the paper, the subscript Q indicates the case of photon-counting measurements at Alice's end.

The marginal state of the $A$ modes is an $M$-mode thermal state,
\begin{align} \label{eq:marginalthermal}
\rho_{\textrm{th},A}
=\trace_B\big(\ketbra{\Psi_{AB}}\big)
=(1-\chi^2)^M\sum_{N=0}^\infty\chi^{2N}\Pi_N
=(1-\chi^2)^M\chi^{2\sum_j a_j^\dagger a_j}\,.
\end{align}
where
\begin{align}\label{eq:PiN}
\Pi_N=\hspace{-5pt}\sum_{{\scriptstyle{\bm{n}_A}\atop\scriptstyle{|\bm{n}_A|=N}}}\hspace{-5pt}\ketbra{\bm{n}_A}
\end{align}
is the projector onto the subspace of $N$ photons, which has dimension
\begin{align}\label{eq:trPiN}
\trace\,[\,\Pi_N]=\binom{N+M-1}{N}\equiv G(N,M)\,.
\end{align}
The marginal state of the $B$ modes at the input to the \LON, $\rho_{\text{th},B}$, is the same as the thermal state~(\ref{eq:marginalthermal}), with $A$s swapped with $B$s in the notation; we omit the designation $A$ or $B$ on $\rho_{\text{th}}$ when there is no risk of confusion.

The $M$-mode thermal state is a product of thermal states for each of the modes,
\begin{align}\label{eq:rhoth}
\rho_{\textrm{th}}
=\bigotimes_{i=1}^{M}\rho_{\textrm{th},i}
=\bigotimes_{i=1}^{M}\bigg((1-\chi^2)\sum_{n=0}^{\infty} \chi^{2n}\ketbra{n}\bigg)\,.
\end{align}
The probability for photon record $\bm{n}_A$ in the $A$ modes is
\begin{align}
P_{\textrm{Q}}(\bm{n}_A)
=\sum_{\bm{n}_B}P_{\textrm{Q}}(\bm{n}_A,\bm{n}_B|\bm{U})
=\bra{\bm{n}_A}\rho_{\textrm{th},A}\ket{\bm{n}_A}
=(1-\chi^2)^M \chi^{2|\bm{n}_A|}\,.
\end{align}

The mean and variance of the number of photons counted in a single $A$ mode have the Bose-Einstein values,
\begin{align}\label{eq:meannin}
\overline n&=\trace\big[\rho_{\text{th}}a^\dagger_i a_i\big]=\frac{\chi^2}{1-\chi^2}\,,\\
(\Delta n)^2&=\overline n(\overline n+1)=\frac{\chi^2}{(1-\chi^2)^2}\,.
\label{eq:variancenin}
\end{align}
Hence the mean photon number counted from all the $A$ modes is $M\overline n=M\chi^2/(1-\chi^2)$.  The case of interest is weak squeezing, characterized as $\chi^2\alt1/\sqrt M$.  For weak squeezing, the mean total count from all the $A$ modes, $\simeq M\chi^2\alt\sqrt M\ll M$, is much smaller than $M$; the number of photons counted almost certainly satisfies $N\alt\sqrt M$, and it is reasonably likely that at most a single photon is counted in any $A$ mode or in any mode at the output of the \LON.  Conditioned on detecting $N$ single photons in the $A$ and $B$ modes, $\bra{\bm{n}_B} \mathcal{U} \ket{\bm{n}_A}$ is equal to the permanent of an $N\times N$ submatrix of the LON transfer matrix~$\bm U$.  This $N \times N$ submatrix is obtained by deleting rows and columns corresponding to no-click, i.e., $n_{A,j}=n_{B,j}=0$.  Therefore, as multiplicative approximation of permanents of complex matrices is \#P hard (very difficult) and following the argument from~\cite{AA}, {\it exact\/} sampling from the joint output probability distribution~(\ref{eq:JointPrRBS}) cannot be efficiently simulated classically, unless the polynomial hierarchy in computational complexity theory collapses to the third level, which is highly implausible.

We can put a bit more flesh on the photon statistics by noting that the probability to detect $N$ photons in the $A$ modes is
\begin{align}\label{eq:PQN}
P_{\textrm{Q}}(N)
=\hspace{-5pt}\sum_{{\scriptstyle{\bm{n}_A}\atop\scriptstyle{|\bm{n}_A|=N}}}\hspace{-5pt}P_{\text{Q}}(\bm{n}_A)
=\trace\,[\,\rho_{\text{th},A}\Pi_N]
=G(N,M)(1-\chi^2)^M\chi^{2N}\,,
\end{align}
whereas the probability to detect $N$ single photons in the $A$ modes is
\begin{align}
\tilde P_{\text{Q}}(N)=\binom{M}{N}(1-\chi^2)^M\chi^{2N}\,.
\end{align}
The asymptotic Gaussian approximation for weak squeezing is
\begin{align}\label{eq:Gaussian}
\tilde P_{\textrm{Q}}(N)\sim\frac{e^{-N^2/M}}{\sqrt{2\pi M\chi^2}}\exp\!\bigg(\!-\frac{(N-M\chi^2)^2}{2M\chi^2}\bigg)\sim e^{-N^2/M}P_{\textrm{Q}}(N)\,;
\end{align}
$P_{\textrm{Q}}(N)$ is a Gaussian with mean and variance both given by $M\chi^2$, and the probability for detecting $N$ single photons in the $A$ modes is smaller than $P_{\textrm{Q}}(N)$ by the factor $e^{-N^2/M}$.  For $N$ within a few standard deviations of the mean of the Gaussian, $e^{-N^2/M}$ is approximately constant with value $e^{-M\chi^4}$.

Another way to characterize the strength of the squeezing is to maximize the probability to detect a particular number of photons, $N_0$, single or not.  The maximum occurs for squeezing strength $\chi^2=N_0/(M+N_0)$.    For $N_0\alt\sqrt M$, the maximum corresponds to weak squeezing.  The Gaussian approximation~(\ref{eq:Gaussian}) looks the same with the replacement $M\chi^2=N_0$, and the factor by which $\tilde P_{\textrm{Q}}(N)$ is subnormalized is $e^{-N_0^2/M}$.

A standard choice for RBS is $N_0=\sqrt M$, which gives $\chi^2=1/(N_0+1)\simeq1/N_0$; then $P_{\textrm{Q}}(N)$ is a Gaussian with mean and variance both given by $M/N_0=\sqrt M$, and $\tilde P_{\textrm{Q}}(N)$ is the same Gaussian subnormalized by a factor $1/e$.  This factor of $1/e$ is the reasonable probability cited above that the photocounts on the $A$ modes are all single counts.  In RBS, the cases of single counts in the $A$ modes lead to single photons being injected into the corresponding input ports of the \LON; that the populated input ports into the \LON\ are selected uniformly and randomly based on the $A$ counts is the source of the appellation randomized boson sampling (RBS).  In this situation, modulo the two conjectures of~\cite{AA}, {\it approximate\/} sampling, i.e., sampling from probability distributions close in the total variation distance to the output probability distribution~(\ref{eq:JointPrRBS}) is also classically~hard~\cite{AA,RBS}.

We consider a RBS protocol that is formulated as a  task distributed between two parties, Alice and Bob, as described more precisely in the following and depicted in Fig.~\ref{fig:RBSsetup}.

\begin{enumerate}
	\item Alice generates and sends an $M\times M$ random unitary matrix $\bm U$ to Bob.  Bob, assumed to have the capability to implement the \LON\ corresponding to any $\bm U$, implements this \LON.
    \item Modes $A$ and $B$ of the SPDC sources are shared between Alice and Bob.
	\item Bob inputs the $B$ modes to his \LON.  Then he detects the number of photons output into each of the $B$ modes and sends Alice the results, which are samples from the output photon-counting probability distribution.
	\item By counting photons in each of the $A$ modes and knowing the outcomes of Bob's measurements, Alice samples from the joint probability distribution of the RBS~problem.
\end{enumerate}
\noindent
In the next section we show that by making a heterodyne measurement on each of the $A$ modes, instead of photocounting, Alice can characterize the \LON\ that Bob has implemented, assuming that it is a \LON\ with photon losses.

\section{Characterization protocol}
\label{sec:characterization}

Here we describe how the characterization protocol works. As shown in Fig.~\ref{fig:RBSsetup}, the output modes of $M$ SPDC sources with the same squeezing parameter $\chi$ are distributed between Alice and Bob. Bob is supposed to implement the LON according to Alice's description and send samples from the output photon-counting probability distribution to Alice.  Alice can make either heterodyne (C) or photon-counting (Q) measurements for each mode.  Using the data from the heterodyne measurements, Alice can characterize, we show, Bob's actual, lossy~\LON; moreover, Alice can compare, we show, Bob's actual~\LON\ to the ideal~\LON\ using a measure of distance between transfer matrices and obtain from this measure an upper bound on the total variation distance between the joint photon-counting probability distribution of the experiment and the ideal RBS probability distribution.  Alice can intersperse the characterization runs with the RBS runs, for which she makes photon-counting measurements, thus checking whether Bob's \LON\ remains the same throughout the course of the RBS experiment.

Throughout this section, we often specialize to weak squeezing, with weak-squeezing results signaled by use of the $\simeq$ in $\chi^2\simeq1/\sqrt M\ll1$.

%%%%%%%%%%%%%%%%%%%%%%%%%%%%%%%%%%%%%%%%%%%%%%%%%%%%%%%%%%%%%%%%%%%%%%%%%%
%%%%%%%%%%%%%%%%%%%%%%%%%%%%%%%%%%%%%%%%%%%%%%%%%%%%%%%%%%%%%%%%%%%%%%%%%%
\subsection{Lossy network and RBS runs}
\label{sec:samplingruns}

In an ideal situation, Bob implements the LON that is requested by Alice, described by a unitary matrix $\bm U$, and sends samples from the output photon-counting probability distribution to Alice.  Alice makes photon-counting measurements whose POVM elements are the multimode number states $\ketbra{\bm{n}_A}$.  Hence, as described earlier, if Bob draws samples from the ideal distribution, Alice samples from the joint probability distribution~(\ref{eq:JointPrRBS}), which can be written as
\begin{align}
\label{PQU}
P_{\textrm{Q}}(\bm{n}_A, \bm{n}_B|\bm U)
= P_{\textrm{Q}}(\bm{n}_A) P_{\text{BS}}(\bm{n}_B|\bm{n}_A,\bm U)
=P(\bm{n}_B|\bm{U})P_{\textrm{Q}}(\bm{n}_A|\bm{n}_B,{\bm U})\,,
\end{align}
where the conditional distribution,
\begin{align}
\label{eq:conditionalPrBS}
P_{\text{BS}}(\bm{n}_B|\bm{n}_A,\bm U)
=\frac{P_{\textrm{Q}}(\bm{n}_A, \bm{n}_B|{\bm U})}{P_{\textrm{Q}}(\bm{n}_A)}
=\big\vert \bra{\bm{n}_B} \mathcal{U} \ket{\bm{n}_A} \big\vert^2
=P_{\textrm{Q}}(\bm{n}_A|\bm{n}_B,{\bm U})\,,
\end{align}
is the boson-sampling probability distribution and
\begin{align}
\label{marginaln}
P_{\textrm{Q}}(\bm{n}_A)=(1-\chi^2)^M \chi^{2N}=P(\bm{n}_B|\bm{U})
\end{align}
are the marginal probability distributions, with $N=|\bm{n}_A|=|\bm{n}_B|$.  Without knowledge of the outcomes of the other party's measurements, Alice's photocounts and Bob's photocounts at the output of the \LON\ are drawn from the $M$-mode thermal state~(\ref{eq:marginalthermal}),
\begin{align}
\bra{\bm{n}}\rho_{\text{th}}\ket{\bm{n}}
=\bra{\bm{n}}\mathcal{U}\,\rho_{\text{th}}\,\mathcal{U}^{\dagger}\ket{\bm{n}}
=(1-\chi^2)^M \chi^{2|\bm{n}|}\,,
\end{align}
which is independent of the \LON.  The marginal thermal state at the input to Bob's \LON\ is transferred to the output because $\mathcal{U}$, being photon-number preserving, commutes with $\rho_{\text{th}}$, i.e.,
\begin{align}\label{eq:UthUdagger}
\mathcal{U}\,\rho_{\text{th}}\,\mathcal{U}^{\dagger}=\rho_{\text{th}}\,.
\end{align}

In the presence of losses, Bob's \LON\ is not unitary, but it can be described uniquely by a {\it transfer matrix\/} $\bm L$ that is defined through the relation between input and output amplitudes of coherent states; i.e., input coherent state $\ket{{\bm\beta}}$ goes to output coherent state $\ket{{\bm\beta}'}=\ket{\bm{\beta L}}$.  The (trace-preserving) quantum operation for the lossy \LON, which we call a quantum process, is thus defined by
\begin{align}
\label{eq:ELbeta}
\mathcal{E}_{\bm{L}}\big(\ketbra{\bm{\beta}}\big)=\ketbra{\bm{\beta L}}\,.
\end{align}
Losses at the input to Bob's \LON\ and inefficiencies in Bob's photodetectors can be incorporated into the transfer matrix $\bm{L}$.  Notice that in general $\bm{L}^\dagger\bm{L}\leq\bm{I}$, so any LON can be thought of as part of a larger LON that is unitary (this approach is developed in App.~\ref{sec:auxiliarymodes}).  One reverts to the ideal, lossless network by setting $\bm{L}=\bm{U}$.

It is important to note the phase freedom in the transfer matrix $\bm{L}$.  We can absorb phases in the transfer matrix into the definitions of output modes of the \LON.  We cannot change the phases of the input modes because those have been set relative to the $A$ modes by making $\chi$ real.  This freedom allows us to choose the phase of one element of each column of $\bm{L}$.  What we find most useful in Sec.~\ref{sec:characterizationruns} is to make the diagonal elements $L_{ii}$ real and~nonnegative.

With lossy {\LON}s, instead of sampling from $P_{\textrm{Q}}(\bm{n}_A, \bm{n}_B|\bm U)$, the samples generated by Alice and Bob are drawn from
\begin{align}\label{eq:lossyJointPrRBS}
P_{\textrm{Q}}(\bm{n}_A, \bm{n}_B|{\bm L})
=\big\langle\bm{n}_A,\bm{n}_B\big\vert\,\mathcal{I}_A\otimes\mathcal{E}_{\bm{L}}\big(\ketbra{\Psi_{AB}}\big)\big\vert\bm{n}_A,\bm{n}_B\big\rangle
=(1-\chi^2)^M\chi^{2|\bm{n}_A|}\big\langle\bm{n}_B\big\vert\mathcal{E}_{\bm{L}}\big(\ketbra{\bm{n}_A}\big)\big\vert\bm{n}_B\big\rangle\,.
\end{align}
Since $\mathcal{E}_{\bm{L}}$ is trace preserving, it is easy to see that
\begin{align}
P_{\textrm{Q}}(\bm{n}_A)
=\sum_{\bm{n}_B}P_{\textrm{Q}}(\bm{n}_A, \bm{n}_B|{\bm L})
=(1-\chi^2)^M\chi^{2|\bm{n}_A|}\trace\big[\mathcal{E}_{\bm{L}}(\ketbra{\bm{n}_A})\big]
=(1-\chi^2)^M\chi^{2|\bm{n}_A|}
\end{align}
is the same as for a unitary \LON---this is obvious because the $A$ modes do not see any losses---and thus that the (conditional) boson-sampling distribution becomes
\begin{align}\label{eq:lossyconditionalPrBS}
P_{\text{BS}}(\bm{n}_B|\bm{n}_A,\bm{L})
=\frac{P_{\textrm{Q}}(\bm{n}_A, \bm{n}_B|{\bm L})}{P_{\textrm{Q}}(\bm{n}_A)}
=\bra{\bm{n}_B} \mathcal{E}_{\bm{L}}\left(\ketbra{\bm{n}_A}\right) \ket{\bm{n}_B}\,.
\end{align}

Not so simple is the conditional probability in the other direction, since the lossy \LON\ reduces the number of counts in the $B$ modes in a nondeterministic way, as one sees from
\begin{align}\label{eq:PnBL1}
P(\bm{n}_B|\bm{L})=\sum_{\bm{n}_A}P_{\textrm{Q}}(\bm{n}_A, \bm{n}_B|{\bm L})
=(1-\chi^2)^M\sum_{N=0}^\infty\chi^{2N}\braket{\bm{n}_B|\mathcal{E}_{\bm{L}}(\Pi_N)|\bm{n}_B}
=\bra{\bm{n}_B}\mathcal{E}_{\bm{L}}(\rho_{\text{th},B})\ket{\bm{n}_B}\,,
\end{align}
where $\Pi_N$ is the projector onto the $B$ subspace with $N$ photons, as in Eq.~(\ref{eq:PiN}).

The questions now are how Alice can characterize the quantum process associated with Bob's LON, how different it is from the ideal \LON, and how far the lossy boson-sampling distribution $P_{\textrm{BS}}(\bm{n}_B|\bm{n}_A,{\bm L})$ departs from the ideal distribution $P_{\textrm{BS}}(\bm{n}_B|\bm{n}_A,{\bm U})$ in total variation distance.

%%%%%%%%%%%%%%%%%%%%%%%%%%%%%%%%%%%%%%%%%%%%%%%%%%%%%%%%%%%%%%%%%%%%%%%%%%
%%%%%%%%%%%%%%%%%%%%%%%%%%%%%%%%%%%%%%%%%%%%%%%%%%%%%%%%%%%%%%%%%%%%%%%%%%
\subsection{Characterization}
\label{sec:characterizationruns}

The aim in the characterization protocol is to determine efficiently the lossy transfer matrix $\bm L$ of the actual \LON\ that Bob has implemented, as best he can, according to Alice's specification of the ideal matrix $\bm U$.   In the next subsection, we show how Alice, with the characterization of $\bm{L}$ in hand, can compare the ideal and lossy {\LON}s and obtain an upper bound on the distance (in total variation) between the joint probability distribution of the experiment and the ideal RBS probability distribution.

We note now the polar decomposition of the transfer matrix,
\begin{align}\label{eq:pdL}
\bm{L}=\sqrt{\bm{L}\bm{L}^\dagger}\,\bm{V}=\bm{V}\sqrt{\bm{L}^\dagger\bm{L}}\,,
\end{align}
where the matrix $\sqrt{\bm{L}\bm{L}^\dagger}$ ($\sqrt{\bm{L}^\dagger\bm{L}}$) represents the losses referred entirely to the input (output) of the \LON, and $\bm{V}$ is the ideal network associated with $\bm{L}$.  The transfer matrix $\bm{L}$ is subunitary, i.e., $\bm{L}\bm{L}^\dagger\le\bm{I}$ ($\bm{L}^\dagger\bm{L}\le\bm{I}$).  Characterizing $\bm{L}$ is a matter of determining the network $\bm{V}$ and the losses $\sqrt{\bm{L}\bm{L}^\dagger}$ (or $\sqrt{\bm{L}^\dagger\bm{L}}$) that go with it. Notice that knowing both loss matrices, $\sqrt{\bm{L}\bm{L}^\dagger}$ and $\sqrt{\bm{L}^\dagger\bm{L}}$, determines $\bm{V}$ and, hence, the entirety of $\bm{L}$.

If we diagonalize the loss matrices with a further unitary $\bm{W}$,
\begin{align}\label{eq:svdLLdagger}
\sqrt{\bm{L}\bm{L}^\dagger}&=\bm{W}\textbf{diag}(t_1,\ldots,t_M)\bm{W}^\dagger\,,\\
\sqrt{\bm{L}^\dagger\bm{L}}&=\bm{V}^\dagger\bm{W}\textbf{diag}(t_1,\ldots,t_M)\bm{W}^\dagger\bm{V}\,,
\label{eq:svdLdaggerL}
\end{align}
the transfer matrix becomes
\begin{align}
\bm{L}=\bm{W}\textbf{diag}(t_1,\ldots,t_M)\bm{W}^\dagger\bm{V}\,.
\label{eq:svdWV}
\end{align}
The elements $t_i$ of the diagonal matrix are the (real and nonnegative) singular values of $\bm{L}$; the subunitarity of $\bm{L}$ implies that $t_i\le1$.
The singular values describe the losses from a set of decoupled, pure-loss modes; physically, they are the transmissivities of a set of $M$ beamsplitters that remove photons from these decoupled modes, as in the loss model developed in App.~\ref{sec:auxiliarymodes}.  The transfer matrix can be thought of as an initial unitary $\bm{W}$ that transforms from the input modes to the pure-loss modes, followed by the pure losses and a final unitary $\bm{W}^\dagger\bm{V}$ that transforms to the output modes.

%%%%%%%%%%%%%%%%%%%%%%%%%%%%%%%%%%%%%%%%%%%%%%%%%%%%%%%%%%%%%%%%%%%%%%%%%%
\subsubsection{Setup for characterization runs}

For the characterization runs, Alice uses heterodyne measurements, whose joint POVM elements are multiples of multimode coherent-state projectors,  $\ketbra{\bm{\alpha}}/\pi^M$. For outcome $\bm{\alpha}$ of the heterodyne measurements, the state at the input of Bob's LON is projected to the multimode coherent state $\ket{\chi\bm{\alpha}^*}$,
\begin{align}\label{eq:alphaonPsi}
\frac{1}{\pi^{M/2}}\braket{\bm{\alpha}|\Psi_{AB}}
=\frac{1}{\pi^{M/2}}\prod_{i=1}^M\braket{\alpha_i|\psi_{AB}}_i
=\frac{(1-\chi^2)^{M/2}}{\pi^{M/2}}e^{-(1-\chi^2)|\bm{\alpha}|^2/2}\ket{\chi\bm{\alpha}^*}\,,
\end{align}
with
\begin{align}
\bm{\alpha}^*=
\begin{pmatrix}
\alpha_1^*&\alpha_2^*&\cdots&\alpha_M^*
\end{pmatrix}\,,
\qquad
|\bm{\alpha}|^2=\bm{\alpha}^*\bm{\alpha}^T=\bm\alpha\bm{\alpha}^{\dagger}=\sum_{i=1}^M|\alpha_i|^2\,.
\end{align}
It is known that one can characterize a \LON\ from the photocount statistics obtained when the network is illuminated with a particular set of coherent states~\cite{Rahimi-Keshari2011,Rahimi-Keshari2013}, so it is not surprising that we can devise a protocol for characterizing the \LON\ from the joint statistics of Alice's heterodyne measurements and Bob's photocount records.

Given heterodyne outcome $\bm{\alpha}$, the state at the output of the \LON\ is the coherent state $\ket{\chi \bm{\alpha}^*\bm{L}}$.  More precisely, we have
\begin{align}
\frac{1}{\pi^M}\big\langle\bm{\alpha}\big\vert\mathcal{I}_A\otimes\mathcal{E}_{\bm{L}}\big(\ketbra{\Psi_{AB}}\big)\big\vert\bm{\alpha}\big\rangle
=\frac{1}{\pi^M}\mathcal{E}_{\bm{L}}\big(\braket{\bm{\alpha}|\Psi_{AB}}\braket{\Psi_{AB}|\bm{\alpha}}\big)
=\frac{(1-\chi^2)^M}{\pi^M}e^{-(1-\chi^2)|\bm{\alpha}|^2}\ketbra{\chi\bm{\alpha}^*\bm{L}}\,.
\end{align}
Further projecting onto the photon-counting outcome $\bm{n}_B$ at Bob's end gives the joint probability for outcomes $\bm\alpha$ and $\bm{n}_B$,
\begin{align}\label{eq:jointPCL}
P_{\textrm{C}}(\bm{\alpha},\bm{n}_B|\bm{L})
=\frac{(1-\chi^2)^M}{\pi^M}e^{-(1-\chi^2)|\bm{\alpha}|^2}\big\vert\braket{\bm{n}_B|\chi\bm{\alpha}^*\bm{L}}\big\vert^2\,,
\end{align}
where
\begin{align}\label{eq:bigPoisson}
\big\vert\braket{\bm{n}_B|\chi \bm{\alpha}^*\bm{L}}\big\vert^2
=\prod_{i=1}^M\big\vert\braket{n_{B,i}|\chi\bm{\alpha}^*\bm{L}_i}\big\vert^2
=\chi^{2|\bm{n}_B|} e^{-\chi^2\bm{\alpha}^*\bm{L}\bm{L}^\dagger\bm{\alpha}^T}
\prod_{i=1}^{M}\frac{\big\vert\bm{\alpha}^*\bm{L}_i\big\vert^{2n_{B,i}}}{n_{B,i}!}\,.
\end{align}
Here
\begin{align}
\bm{\alpha}^*\bm{L}_i=(\bm{\alpha}^*\bm{L})_i=\sum_{j=1}^M\alpha^*_j L_{ji}\,,
\end{align}
with
\begin{align}\label{eq:Li}
\bm{L}_i=\begin{pmatrix}L_{1i}&L_{2i}&\cdots&L_{Mi}\end{pmatrix}^T
\end{align}
being the column vector made from the $i$th column of $\bm{L}$.

We can split up the joint probability as
\begin{align}\label{eq:jointPCLsplitup}
P_{\textrm{C}}(\bm{\alpha},\bm{n}_B|\bm{L})=P_{\textrm{C}}(\bm{\alpha})P_{\textrm{C}}(\bm{n}_B|\bm{\alpha},\bm{L})\,,
\end{align}
where
\begin{align} \label{eq:probCalpha}
P_{\textrm{C}}(\bm{\alpha})=\sum_{\bm{n}_B}P_{\textrm{C}}(\bm{\alpha},\bm{n}_B|\bm{L})
=\frac{(1-\chi^2)^M}{\pi^M}e^{-(1-\chi^2)|\bm{\alpha}|^2}
=\frac{1}{\pi^M}\bra{\bm{\alpha}}\rho_{\textrm{th},A}\ket{\bm{\alpha}}
\end{align}
is the unconditional probability for heterodyne outcome~$\bm{\alpha}$ at Alice's end.  For this unconditional probability, $\bm{\alpha}$ is drawn from the thermal state~(\ref{eq:marginalthermal}); the probability~(\ref{eq:probCalpha}) is the Husimi $Q$ distribution~\cite{Husimi1940} of this thermal state.  The conditional probability for Bob's photocounts, given heterodyne outcome $\bm{\alpha}$,
\begin{align}\label{conditionalPCL}
P_{\textrm{C}}(\bm{n}_B|\bm{\alpha},\bm{L})
=\big\vert\!\braket{\bm{n}_B|\chi \bm{\alpha}^*\bm{L}}\!\big\vert^2\,,
\end{align}
can be efficiently classically computed using Eq.~(\ref{eq:bigPoisson}), as opposed to the distribution~(\ref{eq:conditionalPrBS}), which is \#P hard to compute.

\subsubsection{Bob's unconditional photostatistics}
\label{sec:Bobstats}

We pause briefly to give an account of what can be learned from the unconditional (marginal) photostatistics at Bob's end. The unconditional probability~(\ref{eq:PnBL1}) for photocount record $\bm{n}_B$,
\begin{align}\label{eq:PnBL2}
P(\bm{n}_B|\bm{L})
=\int d^{2M}\!\bm{\alpha}\,P_{\text{C}}(\bm{\alpha},\bm{n}_B|\bm{L})
=\bra{\bm{n}_B}\mathcal{E}_{\bm{L}}(\rho_{\text{th},B})\ket{\bm{n}_B}\,,
\end{align}
is independent of what happens at Alice's end and describes sampling from the output marginal state $\mathcal{E}_{\bm{L}}(\rho_{\text{th},B})$, which is the input marginal thermal state at Bob's end processed through the lossy network.  Since the unconditional photocount distribution $P(\bm{n}_B|\bm{L})$ is independent of any measurement Alice performs, we omit the subscript C in Eq.~(\ref{eq:PnBL2}) and, previously, the subscript Q in Eq.~(\ref{eq:PnBL1}).

The $M$-mode thermal state~(\ref{eq:rhoth}) can be represented by its Glauber-Sudarshan $P$ function~\cite{Glauber,Sudarshan} in the coherent-state expansion
\begin{align}\label{eq:Glauber-Sudarshan}
\rho_{\text{th},B}
=\int d^{2M}\!\bm{\beta}\,
\frac{e^{-\bm{\beta}\bm{\beta}^\dagger/\overline{n}}}{(\pi\overline n)^M}\ketbra{\bm{\beta}}\,.
\end{align}
The $M$-mode marginal state at the output of Bob's \LON\ is
\begin{align}\label{eq:outstateBob}
\mathcal{E}_{\bm{L}}(\rho_{\text{th},B})
=\int d^{2M}\!\bm{\beta}\,
\frac{e^{-\bm{\beta}\bm{\beta}^\dagger/\overline{n}}}{(\pi\overline n)^M}\ketbra{\bm{\beta}\bm{L}}
=\int d^{2M}\!\bm{\gamma}\,
\frac{e^{-\bm{\gamma}(\overline n\bm{L}^\dagger\bm{L})^{-1}\bm{\gamma}^\dagger}}{(\pi\overline n)^M\det(\bm{L}^\dagger\bm{L})}\ketbra{\bm{\gamma}}\,,
\end{align}
where $\bm{\gamma}=\bm{\beta}\bm{L}$.  Notice that $\det(\bm{L}^\dagger\bm{L})=t_1^2\cdots t_M^2$.  This state has, as shown, a Gaussian $P$ function with, in general, correlations between the output modes.  If $\bm{L}$ is unitary, the marginal output state is identical to the input thermal state, as was noted in Eq.~(\ref{eq:UthUdagger}).  Diagonalizing the output loss matrix $\bm{L}^\dagger\bm{L}$, as in Eq.~(\ref{eq:svdLdaggerL}), expresses the output marginal state~(\ref{eq:outstateBob}) in terms of decoupled pure-loss modes, transformed from Bob's output by the matrix $\bm{V}^\dagger\bm{W}$; these decoupled loss modes are each in a thermal state, but of different temperatures and thus with different mean photon numbers, $\overline n t_i^2\le\overline n$.

The lesson here is that the output marginal state is determined by the output loss matrix $\bm{L}^\dagger\bm{L}$ and is independent of the associated lossless transfer matrix $\bm{V}$.  Hence, the unconditional photostatistics at Bob's end---more generally, the unconditional statistics of any measurement at Bob's end---can tell one about $\bm{L}^\dagger\bm{L}$, but provide {\it no\/} information about $\bm{V}$.  Keep in mind that we can collect the unconditional photostatistics from all the runs, both RBS runs and characterization runs.  It should be possible to reconstruct the entire output loss matrix $\bm{L}^\dagger\bm{L}$, up to the phase freedom, from photon-number correlations among the output modes.  First we consider the unconditional photostatistics of each output mode separately, which determine the diagonal elements of $\bm{L}^\dagger\bm{L}$; we comment on the general problem of determining all of $\bm{L}^\dagger\bm{L}$ at the end of this subsection.

The photocount probabilities~$P(\bm{n}_B|\bm{L})$ are proportional, in general, to permanents of positive, Hermitian matrices~\cite{Rahimi-Keshari2015}---these are submatrices of $\bm{L}^\dagger\bm{L}$---and are \#P hard to compute~\cite{Grier}.   Nonetheless, as discussed in~\cite{Rahimi-Keshari2015}, it is easy to construct an efficient classical protocol, based on sampling from the well-behaved positive $P$ function in Eq.~(\ref{eq:outstateBob}), for sampling from $P(\bm{n}_B|\bm{L})$; moreover, Bob's photocount statistics can be used for characterization of $\bm{L}^\dagger\bm{L}$.  These considerations indicate that hardness of computing a distribution, hardness of sampling from it, and characterization based on sampling from it are independent.

To get the marginal state in mode $i$ at Bob's end, it is easiest to work with the normally ordered characteristic function of the $M$-mode output state~(\ref{eq:outstateBob}), which is the Fourier transform of the $P$ function and is given by the expectation value of the normally ordered displacement operator,
\begin{equation}\label{eq:Phitotal}
\Phi\big(\bm\xi|\mathcal{E}_{\bm{L}}(\rho_{\text{th}})\big)
=\trace\big[e^{-\bm{b}\bm{\xi}^\dagger}\mathcal{E}_{\bm{L}}(\rho_{\text{th}})e^{\bm{\xi}\bm{b}^\dagger}\big]
=e^{-\bm\xi(\overline n\bm{L}^\dagger\bm{L})\bm\xi^\dagger}\,.
\end{equation}
Here
\begin{align}\label{eq:rowb}
\bm{b}=\begin{pmatrix}b_1&b_2&\cdots&b_M\end{pmatrix}
\end{align}
is the row vector of annihilation operators.  Marginalization to the normally ordered characteristic function of output mode~$i$ is achieved by setting $\xi_j=0$ for $j\ne i$,
\begin{equation}\label{eq:Phii}
\Phi\big(\xi_i|\mathcal{E}_{\bm{L}}(\rho_{\text{th}})\big)=e^{-\overline n\ell_i{\,^2}|\xi_i|^2}\,,
\end{equation}
where
\begin{align}\label{eq:elli}
\ell_i^{\,2}
=(\bm{L}^\dagger\bm{L})_{ii}
=\sum_{j=1}^M|L_{ji}|^2
=\bm{L}_i^\dagger\bm{L}_i
\le1
\end{align}
is the $i$th diagonal component of $\bm{L}^\dagger\bm{L}$ or, equivalently, the squared length of the column vector $\bm{L}_i$.  The characteristic function~(\ref{eq:Phii}) is that of a thermal state with mean photon number
\begin{align}
\overline{n}_i=\overline n\ell_i^{\,2}=\frac{\chi^2\ell_i^{\,2}}{1-\chi^2}\simeq\chi^2\ell_i^{\,2}\,.
\end{align}
The final, approximate expression here and in similar circumstances below holds for weak squeezing and is good to first order in $\chi^2\simeq1/\sqrt M$.  The probability to count $n_i$ photons from output mode~$i$ is thus
\begin{align}\label{eq:PniL}
P(n_i|\bm{L})
=\frac{1}{1+\overline n_i}\!\left(\frac{\overline n_i}{1+\overline n_i}\right)^{n_i}
=\frac{1-\chi^2}{1-\chi^2(1-\ell_i^{\,2})}\!\left(\frac{\chi^2\ell_i^{\,2}}{1-\chi^2(1-\ell_i^{\,2})}\right)^{n_i}\,.
\end{align}
For weak squeezing, this distribution simplifies to $P(0_i|\bm{L})\simeq1-\chi^2\ell_i^{\,2}$ and $P(1_i|\bm{L})\simeq\chi^2\ell_i^{\,2}$.

By using all the data reported by Bob, from the RBS runs and the characterization runs, Alice can estimate $\ell_i^{\,2}$ for each output mode.  With these estimates in hand, Alice can determine a measure of how far $\bm{L}$ departs from being unitary, i.e.,
\begin{align}
E(\bm L)
=\frac{1}{M}\Vert\bm{I}-\bm{L}^\dagger\bm{L}\Vert_1=1-\frac{1}{M}\trace[\bm{L}^\dagger\bm{L}]=1-\frac{1}{M}\sum_{i=1}^M\ell_i^{\,2}\,,
\end{align}
where the 1-norm (also called the {\it trace norm}) is defined by $\Vert A\Vert_1=\trace\big[\sqrt{A^{\dagger}A}\,\big]$.  This quantity, which satisfies $0\leq E(\bm L) \leq 1$, can be interpreted as the average loss per mode and is also given by the singular values $t_i$, i.e.,
\begin{align}
E(\bm{L})=1-\frac{1}{M}\sum_{i=1}^M t_i^2\,.
\end{align}

As noted above, it should be possible to reconstruct the entire output loss matrix $\bm{L}^\dagger\bm{L}$ from photon-number correlations among the output modes.  The second-order correlations provide information about the magnitudes of the off-diagonal components of $\bm{L}^\dagger\bm{L}$.  Third-order correlations provide information about the phases of the off-diagonal components of $\bm{L}^\dagger\bm{L}$, but this information is not complete information.  Moreover, the relevant parts of the second- and third-order correlations are proportional to $\bar n^2\propto\chi^4$ and $\bar n^3\propto\chi^6$, thus making it difficult to read out the relevant information from the statistics.

A more important question is what kind of \LON\ characterization can be achieved using the joint photocount statistics in the RBS runs.  This question, suggested to us by a comment from a referee, we had not considered to be within the purview of this paper, but it can be addressed and answered using the techniques developed in App.~\ref{sec:hetstats} for determining what kind of characterization can be achieved when the input two-mode squeezed state is replaced by the classical-classical state of Eq.~(\ref{eq:rhoCC}).  The answer, buried in the final subsection, App.~\ref{sec:RBSonly}, of that appendix, deserves more than an afterthought, so we intend to exhume the analysis and give it the attention it deserves in a separate publication.

For now, however, we set aside all questions of characterization using photostatistics, this not being the point of our paper, and turn to analyzing the heterodyne characterization runs, which can determine the entirety of the transfer matrix~$\bm L$.

%%%%%%%%%%%%%%%%%%%%%%%%%%%%%%%%%%%%%%%%%%%%%%%%%%%%%%%%%%%%%%%%%%%%%%%%%%
\subsubsection{Characterization runs}
\label{sec:characterizationdetail}

The conditional probability for heterodyne outcomes,
\begin{align}\label{eq:PalphanBL}
P_{\textrm{C}}(\bm{\alpha}|\bm{n}_B,\bm{L})
=\frac{P_{\textrm{C}}(\bm{\alpha},\bm{n}_B|\bm{L})}{P(\bm{n}_B|\bm{L})}
=\frac{P_{\textrm{C}}(\bm{\alpha})}{P(\bm{n}_B|\bm{L})}\big\vert\!\braket{\bm{n}_B|\chi \bm{\alpha}^*\bm{L}}\!\big\vert^2\,,
\end{align}
obtained from combining Eqs.~(\ref{eq:jointPCL}) and~(\ref{eq:PnBL2}), is the Husismi $Q$ distribution of the state for Alice's modes, conditioned on a particular photocount record at Bob's end.  We now show how, by sampling from this conditional distribution, Alice can characterize fully the transfer matrix $\bm L$ associated with Bob's \LON.  To see clearly what information is available from the heterodyne statistics, we notice that $|\bm{\alpha}^*\bm{L}_i\big\vert^2=(\bm{\alpha}^*\bm{L}_i)(\bm{L}_i^\dagger\bm{\alpha}^T)=\bm{\alpha}^*(\bm{L}_i\bm{L}_i^\dagger)\bm{\alpha}^T
=\bm{L}_i^\dagger\bm{\alpha}^T\bm{\alpha}^*\bm{L}_i$ and stress that $\bm{L}_i\bm{L}_i^\dagger$ is the matrix multiplication, i.e., outer product, of the column vector $\bm{L}_i$ and the row vector $\bm{L}_i^\dagger$.  This in mind, we write the crucial part of the distribution~(\ref{eq:PalphanBL}) in a form subtly different from Eq.~(\ref{eq:bigPoisson}),
\begin{align}\label{eq:bigPoisson2}
\big\vert\braket{\bm{n}_B|\chi \bm{\alpha}^*\bm{L}}\big\vert^2
=\chi^{2|\bm{n}_B|}
\exp\!\bigg[\!-\chi^2\bm{\alpha}^*\bigg(\sum_{i=1}^M\bm{L}_i\bm{L}_i^\dagger\bigg)\bm{\alpha}^T\bigg]
\prod_{i=1}^{M}\frac{\big\vert\bm{\alpha}^*\bm{L}_i\bm{L}_i^\dagger\bm{\alpha}^T\big\vert^{n_{B,i}}}{n_{B,i}!}\,,
\end{align}
which makes clear that what we can hope to characterize from the heterodyne statistics are the matrix elements of the outer products $\bm{L}_i\bm{L}_i^\dagger$.  A little thought shows that determining all these outer products is sufficient to determine $\bm{L}$, with respect to the reference phases of Alice's modes, which make $\chi$ real for every mode, and within the phase freedom of the transfer matrix already discussed.  Appendix~\ref{sec:hetstats} gives a general account of where and how information about $\bm{L}$ is stored in the conditional heterodyne distribution of Eq.~(\ref{eq:PalphanBL}).

To do this characterization, however, we need not consider, nor should we consider the conditional heterodyne distribution for all of Bob's photocount records; indeed, to have an efficient characterization procedure, we should consider the most likely photocount records.  In particular, we focus on the situation where the only condition is that the $i$th mode on Bob's side reports no photocounts.  The joint probability $P_{\textrm{C}}(\bm{\alpha},0_i|\bm{L})$ for heterodyne outcomes $\bm\alpha$ and for no photocounts in the $i$th mode is obtained by summing $P_{\textrm{C}}(\bm{\alpha},\bm{n}_B|\bm{L})$ over all possible photocounts for modes other than $i$ and setting $n_{B,i}=0$ and thus is given by
\begin{align}\label{eq:PCalphai}
\begin{split}
P_{\textrm{C}}(\bm{\alpha},0_i|\bm{L})
&=\frac{(1-\chi^2)^M}{\pi^M}\;e^{-(1-\chi^2)|\bm{\alpha}|^2}\big\vert\braket{0|\chi\bm{\alpha}^*\bm{L}_i}\big\vert^2\\
&=\frac{(1-\chi^2)^M}{\pi^M}\;e^{-(1-\chi^2)|\bm{\alpha}|^2}e^{-\chi^2|\bm{\alpha}^*\bm{L}_i|^2}\\
&=\frac{(1-\chi^2)^M}{\pi^M}\;e^{-\bm{\alpha}^*\ourmatrix_i\bm{\alpha}^T}\,,
\end{split}
\end{align}
where $\ourmatrix_i$ is the positive (Hermitian) matrix
\begin{align}\label{eq:ourMi}
\ourmatrix_i=(1-\chi^2)\bm{I}+\chi^2\bm{L}_i\bm{L}_i^\dagger\,.
\end{align}
It is productive to write
\begin{align}
\bm{L}_i=\ell_i\hat{\bm{L}}_i\,,
\end{align}
where $\ell_i$ is the length of $\bm{L}_i$, given by Eq.~(\ref{eq:elli}), and $\hat{\bm{L}}_i$ is a (complex) unit vector.  This gives $\bm{L}_i\bm{L}_i^\dagger=\ell_i^{\,2}\,\hat{\bm{L}}_i\hat{\bm{L}}_i^\dagger$ and allows us to write $\ourmatrix_i$ in the diagonalized form,
\begin{align}\label{eq:ourMi2}
\ourmatrix_i
=(1-\chi^2)\big(\bm{I}-\hat{\bm{L}}_i\hat{\bm{L}}_i^\dagger\big)
+\big[1-\chi^2(1-\ell_i^{\,2})\big]\hat{\bm{L}}_i\hat{\bm{L}}_i^\dagger\,,
\end{align}
from which it is trivial to see that
\begin{align}
\det\ourmatrix_i=(1-\chi^2)^{M-1}\big[1-\chi^2(1-\ell_i^{\,2})\big]\,.
\end{align}

The unconditional probability for the $i$th mode to have no photocounts is
\begin{align}\label{eq:P0i}
P(0_i|\bm{L})
=\int d^{2M}\!\bm{\alpha}\,P_{\textrm{C}}(\bm{\alpha},0_i|\bm{L})
=\frac{(1-\chi^2)^M}{\det\bm{S}_i}
=\frac{1-\chi^2}{1-\chi^2(1-\ell_i^{\,2})}
\simeq1-\chi^2\ell_i^{\,2},
\end{align}
which follows immediately from the integral or from the thermal photocount distribution for mode~$i$ in Eq.~(\ref{eq:PniL}).  The joint distribution~(\ref{eq:PCalphai}) and the no-count distribution~(\ref{eq:P0i}) together give rise to the conditional distribution for heterodyne outcomes $\bm{\alpha}$, given no counts in the $i$th output mode of the \LON:
\begin{align}\label{eq:PCalpha0L}
P_{\textrm{C}}(\bm{\alpha}|0_i,\bm{L})
=\frac{P_{\textrm{C}}(\bm{\alpha},0_i|\bm{L})}{P(0_i|\bm{L})}
=\frac{\det\ourmatrix_i}{\pi^M}
\;e^{-\bm{\alpha}^*\ourmatrix_i\bm{\alpha}^T}\,.
\end{align}
This conditional distribution is a normalized, zero-mean Gaussian function of $\bm{\alpha}$---it is the Husimi $Q$ distribution of the state of Alice's modes, conditioned on no counts in Bob's mode~$i$---which depends only on $\chi$ and $\bm{L}_i$ and is characterized completely by the covariance matrix (because this is a zero-mean Gaussian, this is also the correlation matrix),
\begin{align}\label{eq:alphacorrelationi}
\braket{\bm{\alpha}^T\bm{\alpha}^*}_i=\int d^{2M}\!\bm{\alpha}\,\bm{\alpha}^T\bm{\alpha}^*P(\bm{\alpha}|0_i,\bm{L})=\ourmatrix_i^{-1}\,.
\end{align}
This general result holds for any zero-mean Gaussian; in our case, the inverse is easily seen to be
\begin{align}\label{eq:ourMiinverse}
\ourmatrix_i^{-1}
=\frac{1}{1-\chi^2}\big(\bm{I}-\hat{\bm{L}}_i\hat{\bm{L}}_i^\dagger\big)
+\frac{1}{1-\chi^2(1-\ell_i^{\,2})}\hat{\bm{L}}_i\hat{\bm{L}}_i^\dagger
=\frac{1}{1-\chi^2}\bigg(\bm{I}-\frac{\chi^2\bm{L}_i\bm{L}_i^\dagger}{1-\chi^2(1-\ell_i^{\,2})}\bigg)\,.
\end{align}
The variance along the complex direction (mode) $\hat{\bm{L}}_i$ is
\begin{align}
\cvar^2=\frac{1}{1-\chi^2(1-\ell_i^{\,2})}\simeq1+\chi^2(1-\ell_i^{\,2})\,,
\end{align}
and the variance along the $M-1$ complex directions (modes) orthogonal to $\hat{\bm{L}}_i$ is
\begin{align}
\cvar_0^2=\frac{1}{1-\chi^2}\simeq1+\chi^2\,.
\end{align}
Notice that $\cvar_0^2\ge\cvar^2\ge1$.  For a lossless \LON, for which $\bm{L}$ is unitary, $\ell_i=1$ and thus $\cvar=1$.  Written with less abstraction, the correlation matrix~(\ref{eq:alphacorrelationi}) has the form
\begin{align}\label{eq:alphajki}
\braket{\alpha_j\alpha_k^*}_i=(S_i^{-1})_{jk}
=\frac{1}{1-\chi^2}\bigg(\delta_{jk}-\frac{\chi^2 L_{ji}L_{ki}^*}{1-\chi^2(1-\ell_i^{\,2})}\bigg)
\simeq\delta_{jk}(1+\chi^2)-\chi^2 L_{ji}L_{ki}^*\,.
\end{align}

A nice way to think of the distribution~(\ref{eq:PCalpha0L}) is in terms of a covariance ellipsoid defined as the unit level surface of the quadratic form $\bm{\alpha}^*\ourmatrix_i\bm{\alpha}^T$; this ellipsoid is nearly a sphere with radius $\cvar_0$, except that along the (complex) axis $\hat{\bm{L}}_i$, it has radius $\cvar\le\cvar_0$.  The covariance ellipsoid deviates from a sphere as much as possible when $\bm{L}$ is unitary. As the distribution~(\ref{eq:PCalpha0L}) is Gaussian, it can be efficiently characterized from the outcomes of the heterodyne measurements; i.e., the heterodyne outcomes can efficiently estimate the orientation of the covariance ellipsoid and its radius along the one special axis, thus determining the $i$th column of $\bm L$.  Since the covariance ellipsoid determines and is determined by the correlation matrix of the distribution---i.e., by the covariance matrix---what we do is to estimate the second-moment correlations from the heterodyne outcomes and use these second moments to estimate the vector $\bm{L}_i$.

For the $i$th mode, Alice needs to estimate the $M$ moments that have $k=i$,
\begin{align}
\braket{\alpha_j\alpha_i^*}_i=\frac{1}{1-\chi^2}\bigg(\delta_{ji}-\frac{\chi^2 L_{ji}L_{ii}^*}{1-\chi^2(1-\ell_i^{\,2})}\bigg)
\end{align}
(thus, overall, for all modes~$i$, she not surprisingly estimates $M^2$ moments); particularly, we have for $j=i$ and $j\ne i$,
\begin{align}
\label{moment_ii}
\braket{|\alpha_i|^2}_i
&=\frac{1}{1-\chi^2}\bigg(1-\frac{\chi^2\vert L_{ii}\vert^2}{1-\chi^2(1-\ell_i^{\,2})}\bigg)
\simeq1+\chi^2(1-|L_{ii}|^2),\\
\braket{\alpha_j\alpha_i^*}_i&=-\frac{\chi^2 L_{ji}L_{ii}^*}{(1-\chi^2)\big[1-\chi^2(1-\ell_i^{\,2})\big]}\simeq-\chi^2 L_{ji}L_{ii}^*,\quad j\ne i\,.
\label{moment_ji}
\end{align}
These moments confirm our earlier discussion of the phase freedom in the transfer matrix: once the diagonal elements $L_{ii}$ are chosen to be real and nonnegative, a choice we make henceforth, the phases of the off-diagonal elements of $\bm{L}$ are determined by the heterodyne statistics.  The variance $\braket{|\alpha_i|^2}_i$ gives the diagonal element
\begin{align}\label{eq:Lii}
L_{ii}=\sqrt{1-\frac{\braket{|\alpha_i|^2}_i-1}{\chi^2}}\,,
\end{align}
and the cross-moments $\braket{\alpha_j\alpha_i^*}_i$ give the off-diagonal elements,
\begin{align}\label{eq:Lji}
L_{ji}&=-\frac{\braket{\alpha_j\alpha_i^*}_i}{\chi^2L_{ii}},\quad j\ne i\,.
\end{align}
The form of the estimates~(\ref{eq:Lii}) and (\ref{eq:Lji}) assumes weak squeezing.

From the heterodyne outcomes of $T_i$ characterization runs for which Bob's output mode~$i$ has no counts, we can determine the elements of the $i$th covariance matrix with uncertainty $\sim1/\sqrt{T_i}$, which means that we can determine the elements of the $i$th column of $\bm{L}$ with uncertainty $\sim1/\chi^2\sqrt{T_i}$.  In a total of $T$ characterization runs, $T_i$ is a binomial random variable governed by the probability $p_i=P(0_i|\bm{L})$ of Eq.~(\ref{eq:P0i}) and thus having mean $\langle T_i\rangle=Tp_i\simeq T(1-\chi^2\ell_i^{\,2})$ and variance $Tp_i(1-p_i)\simeq T\chi^2\ell_i^{\,2}$.  Thus, from $T$ characterization runs, we have, for each output mode, a subset of nearly $T$ runs that have no count in the chosen output mode.  This means that we can estimate all the elements of the transfer matrix with uncertainty $\delta\sim1/\chi^2\sqrt T\simeq\sqrt{M/T}$, so the required number of characterization runs,
\begin{align}\label{eq:T}
T\sim M/\delta^2\,,
\end{align}
grows linearly with problem size.  It is worth emphasizing that this result comes from the ability of the characterization runs to extract complete information about the transfer matrix from heterodyne second moments, i.e., from the first-order coherence of interfering field complex amplitudes in Alice's modes.  Were we to need higher-order moments, involving higher-order coherences, the required number of characterization runs would be a polynomial of higher order than linear.

Another point worth stressing is the difference between the Alice's conditional heterodyne statistics and Bob's unconditional photocount statistics.  The conditional heterodyne statistics of the characterization runs provide information about all the columns $\bm{L}_i$ of the transfer matrix $\bm{L}=\bm{V}\sqrt{\bm{L}^\dagger\bm{L}}$ and thus provide complete information about both the lossless matrix $\bm{V}$ and the output loss matrix $\bm{L}^\dagger\bm{L}$.  In contrast, measurements on Bob's marginal output state only provide information about the output loss matrix, whose matrix elements,
\begin{align}
(\bm{L}^\dagger\bm{L})_{ij}=\sum_{k=1}^M L^*_{ki}L_{kj}=\bm{L}_i^\dagger\bm{L}_j\,,
\end{align}
are the inner products of the column vectors.  These inner products are invariant under unitary transformations, so cannot provide any information about $\bm{V}$.  Physically, this is the statement that Alice's conditional heterodyne statistics are sensitive to interference effects within the \LON\ that Bob's marginal output state does not know about.

\subsection{Comparing probability distributions}
\label{sec:comparing}

Having characterized Bob's LON, Alice can determine whether Bob's RBS samples are drawn from a probability distribution that is close enough to the desired, ideal probability distribution.  This determination is based on a comparison of the ideal and lossy joint distributions, $P_{\textrm{Q}}(\bm{n}_A, \bm{n}_B|\bm{U})$ and $P_{\textrm{Q}}(\bm{n}_A, \bm{n}_B|\bm{L})$.  In developing this comparison, we come upon a measure of the distance between the ideal and lossy {\LON}s and show that this measure provides a sufficient condition to determine whether the samples of the joint experiment by Alice and Bob are drawn from a probability distribution that is close enough to the desired, ideal probability distribution.

We begin by reviewing measures for comparing two probability distributions or two quantum states (density operators)~\cite{Fuchs}.  There are two well-known classical measures for comparing two probability distributions, $P$ and $Q$, which are defined on the same sample space~\cite{Fuchs}, labeled by outcomes $x$: the {\it total variation distance\/} (Kolmogorov distance or $l_1$ distance),
\begin{align}\label{eq:tvd}
D(P,Q)=\frac12\sum_x\vert P(x)-Q(x)\vert\,,
\end{align}
and the {\it classical fidelity\/} (Bhattacharyya overlap),
\begin{align}\label{eq:cfid}
F(P,Q)=\sum_x\sqrt{P(x)\,Q(x)}\,.
\end{align}
The corresponding measures for two quantum states, $\rho$ and $\sigma$, are the {\it trace distance},
\begin{align}\label{eq:td}
D(\rho,\sigma)=\frac12\Vert\rho-\sigma\Vert_1=\frac12\trace\,|\rho-\sigma|=\max_{\{E_x\}}D(P,Q)\,,
\end{align}
and the fidelity,
\begin{align}\label{eq:qf}
F(\rho,\sigma)=\trace\Big[\big(\sqrt{\rho}\sigma\sqrt{\rho}\big)^{1/2}\Big]=\min_{\{E_x\}}F(P,Q)\,.
\end{align}
The final expressions relate the quantum measures between states $\rho$ and $\sigma$ to the classical measures between probability distributions $P$ and $Q$ obtained from measurements on those states; in these expressions, $\{E_x\}$ denotes a POVM, and $P(x)=\trace[\rho E_x]$ and $Q(x)=\trace[\sigma E_x]$.  All these measures are real and symmetric.  If $\rho=\ketbra{\psi}$ is a pure state, $F(\rho,\sigma)=\bra\psi\sigma\ket\psi^{1/2}$.  The classical and quantum measures obey an important set of inequalities~\cite{Fuchs}, which are summarized in the following array:
\begin{align}\label{eq:inequalities}
	\begin{array}{ccccc}
1-F(P,Q)                        &\le        &D(P,Q)                         &\le        &\sqrt{1-F^2(P,Q)}\\
\rotatebox[origin=c]{90}{$\ge$} &           &\rotatebox[origin=c]{90}{$\ge$}&           &\rotatebox[origin=c]{90}{$\ge$}\\
1-F(\rho,\sigma)                &\le        &D(\rho,\sigma)                 &\le        &\sqrt{1-F^2(\rho,\sigma)}
	\end{array}
\,.
\end{align}

The total variation distance between distributions $P$ and $Q$ has an operational interpretation as the probability of error in distinguishing the two distributions: the probability of error given a single draw from the distributions is $\frac12[1-D(P,Q)]$~\cite{Fuchs}.  Through the connection to measurements, the trace distance is directly related to the probability of error in distinguishing density operators $\rho$ and $\sigma$: the probability of error, given an optimal measurement, is $\frac12[1-D(\rho,\sigma)]$; this is known as the {\it Helstrom bound\/}~\cite{Helstrom}.

In the context we are considering, we want to compare the RBS distributions~(\ref{eq:conditionalPrBS}) and~(\ref{eq:lossyconditionalPrBS}) for the ideal and lossy networks using both the total variation distance and the classical fidelity,
\begin{align}\label{eq:tvdBS}
\tvdBS&=\frac12\sum_{\bm{n}_B}\big\vert P_{\textrm{BS}}(\bm{n}_B|\bm{n}_A,\bm{U})-P_{\textrm{BS}}(\bm{n}_B|\bm{n}_A,\bm{L})\big\vert\,,\\
\bcBS&=\sum_{\bm{n}_B}\sqrt{P_{\textrm{BS}}(\bm{n}_B|\bm{n}_A,\bm{U})P_{\textrm{BS}}(\bm{n}_B|\bm{n}_A,\bm{L})}\,.
\label{eq:bcBS}
\end{align}
What we are really interested in---and what we bound in App.~\ref{sec:thefidelitybound}---are the averages of these two quantities over $P_{\textrm{Q}}(\bm{n}_A)$, which are the corresponding quantities for the joint distributions~(\ref{eq:JointPrRBS}) and (\ref{eq:lossyJointPrRBS}),
\begin{align}\label{eq:tvd2}
\sum_{\bm{n}_A}P_{\textrm{Q}}(\bm{n}_A)\tvdBS
&=\frac12\sum_{\bm{n}_A,\bm{n}_B}\big\vert P_{\textrm{Q}}(\bm{n}_A, \bm{n}_B|\bm{U})-P_{\textrm{Q}}(\bm{n}_A, \bm{n}_B|\bm{L})\big\vert
=\tvd,\\
\sum_{\bm{n}_A}P_{\textrm{Q}}(\bm{n}_A)\bcBS
&=\sum_{\bm{n}_A,\bm{n}_B}\sqrt{P_{\textrm{Q}}(\bm{n}_A,\bm{n}_B|\bm{U}) P_{\textrm{Q}}(\bm{n}_A, \bm{n}_B|\bm{L})}
=\bc.
\label{eq:bc}
\end{align}
In the following we focus on two related bounds on these quantities, which involve the quantum measures for Alice and Bob's joint states after processing through the ideal and lossy \LON s.  In App.~\ref{sec:fidelityboundsfixedN}, we formulate fidelity bounds that instead of averaging over all input photon records $\bm{n}_A$, average over records with a fixed total photon number.

To state the bounds compactly, we need to introduce more efficient notation for the joint quantum states.  For this purpose, let $\rho_{AB}=\ketbra{\Psi_{AB}}$ denote the joint input state, let
\begin{align}
\ket{\Psi_{AB|\bm{U}}}=\mathcal{I}_A\otimes\mathcal{U}\ket{\Psi_{AB}}
\end{align}
be the joint (pure) state after the ideal \LON, and let
\begin{align}
\rho_{AB|\bm{U}}&=\ketbra{\Psi_{AB|\bm{U}}}
=\mathcal{I}_A\otimes\mathcal E_{\bm U}(\rho_{AB})\,,\\
\rho_{AB|\bm{L}}&=\mathcal{I}_A\otimes\mathcal E_{\bm L}(\rho_{AB})
\end{align}
be the joint density operators after the ideal and lossy \LON s.

The first bound we consider involves the fidelities,
\begin{align}\label{eq:fidelitybound}
\begin{split}
\bc\ge\qf
=\sqrt{\big\langle\Psi_{AB|\bm{U}}\big|\rho_{AB|\bm{L}}\big|\Psi_{AB|\bm{U}}\big\rangle}
=\frac{(1-\chi^2)^M}{\big\vert\det(\bm{I}-\chi^2\bm{L}\,\bm{U}^\dagger)\big\vert}\,.
\end{split}
\end{align}
We explicitly derive this bound in App.~\ref{sec:thefidelitybound}, not relying on the inequality implicit in Eq.~(\ref{eq:qf}), because this derivation identifies the conditions for saturating the bound and provides the explicit form for the quantum fidelity shown here.  The entirety of App.~\ref{sec:fidelitylowerbound} is as instructive as the main text, both in the derivation of fidelity bounds and in the formulation of a lossy \LON\ in terms of auxiliary modes into which lost photons leak, but telling that story here would interrupt the flow too much.  Appendix~\ref{sec:fidelityboundsfixedN} builds on the apparatus of App.~\ref{sec:thefidelitybound} with a brief investigation of other fidelity measures, which involve input number states rather than thermal states, and how these are related to the transfer matrices.

The second bound weaves through the array~(\ref{eq:inequalities}) to put a bound on the total variation distance,
\begin{align}\label{eq:tvdbound}
\tvd\le\qtd\le\sqrt{1-\big[\qf\big]^2}
=\sqrt{1-\frac{(1-\chi^2)^{2M}}{\big\vert\det(\bm{I}-\chi^2\bm{L}\,\bm{U}^\dagger)\big\vert^2}}\,.
\end{align}
As expected, if $\bm{L}=\bm{U}$, the fidelity bound~(\ref{eq:fidelitybound}) goes to unity, and the bound~(\ref{eq:tvdbound}) on total variation distance goes to zero, meaning that the probabilities and density operators are the same.

It is useful to relate the measurement-independent distance measures we are considering to measures for comparing quantum processes.  The trace distance in Eq.~(\ref{eq:tvdbound}) can be written in a form,
\begin{align}\label{eq:qtd}
\qtd
=\frac12\big\Vert\rho_{AB|\bm{U}}-\rho_{AB|\bm{L}}\big\Vert_1
=\frac12\big\Vert\mathcal{I}_A\otimes\big(\mathcal{E}_{\bm{U}}-\mathcal{E}_{\bm{L}}\big)(\rho_{AB})\big\Vert_1\,,
\end{align}
which invites one to think of it as a species of diamond norm~\cite{Aharonov} between the ideal and lossy quantum networks.  The diamond norm, however, involves a maximization over input states $\rho_{AB}$.  In infinite-dimensional Hilbert spaces, the maximum is generally meaningless, so instead one contemplates maximizing over states with a constraint on, say, the mean energy of the state; measures of this sort have been explored, with the constraint on the energy of the system~$B$~\cite{Shirokov,Winter} or on the joint system $AB$~\cite{Pirandola}, and are called energy-constrained diamond norms. The trace distance~(\ref{eq:qtd}) does not involve any state maximization, but rather probes the \LON s with the entangled multimode squeezed state $\ket{\Psi_{AB}}$, with the squeezing parameter $\chi$ acting as an energy (or photon-number) scale on which $\ket{\Psi_{AB}}$ probes the \LON s.  Hence, for an energy constraint that is set by the squeezing parameter, the trace distance~(\ref{eq:qtd}) is a lower bound for the energy-constrained diamond norm.  Because $\ket{\Psi_{AB}}$ is symmetric between the identical sets of modes, $A$ and $B$, the energy scale set by the squeezing parameter can be regarded either as joint property of all the modes or as a property of the $B$ modes.

More directly relevant to a comparison of \LON s as quantum processes than a diamond norm is the quantum fidelity that provides the bound, whose square we manipulate in the following way:
\begin{align}
\begin{split}
\big[\qf\big]^2
&=\big\langle\Psi_{AB}\big|
\mathcal{I}_A\otimes\mathcal{U}^\dagger\mathcal{I}_A\otimes\mathcal{E}_{\bm L}\big(\rho_{AB}\big)\mathcal{I}_A\otimes\mathcal{U}\big|\Psi_{AB}\big\rangle\\
&=\big\langle\Psi_{AB}\big|
\mathcal{I}_A\otimes\big(\mathcal{E}^{-1}_{\bm{U}}\circ\mathcal{E}_{\bm L}\big)\big(\rho_{AB}\big)\big|\Psi_{AB}\big\rangle\\
&=\fe\,.
\end{split}
\end{align}
The square is the \textit{entanglement fidelity\/}~\cite{Schumacher} of the marginal thermal state of the $B$ modes subjected to a quantum process that is the lossy \LON\ followed by the inverse of the ideal \LON.  Despite its appearance, the entanglement fidelity depends only on the state of the $B$ modes, here the thermal state $\rho_{\text{th},B}$, and the quantum process, here $\mathcal{E}^{-1}_{\bm{U}}\circ\mathcal{E}_{\bm L}$, and is independent of the particular purification of the marginal state, here the two-mode squeezed state $\ket{\Psi_{AB}}$.  We suggest that the entanglement fidelity,
\begin{align}
\fe
=\big\langle\Psi_{AB|\bm{U}}\big|\rho_{AB|\bm{L}}\big|\Psi_{AB|\bm{U}}\big\rangle
=\frac{(1-\chi^2)^{2M}}{\big\vert\det(\bm{I}-\chi^2\bm{L}\,\bm{U}^\dagger)\big\vert^2}\,,
\end{align}
is a good, measurement-independent measure for comparing a lossy \LON\ $\bm{L}$ with an ideal, lossless network $\bm{U}$.  The entanglement fidelity does depend on the marginal thermal state $\rho_{\text{th},B}$ input to the \LON.  In App.~\ref{sec:fidelityboundsfixedN} we develop related fidelity bounds that instead of using the canonical ensemble of the thermal state, are based on microcanonical ensembles, i.e., mixed states spread uniformly over all states with the same total number of photons.

\begin{figure}[t]
	\centering
	\includegraphics[scale=.45]{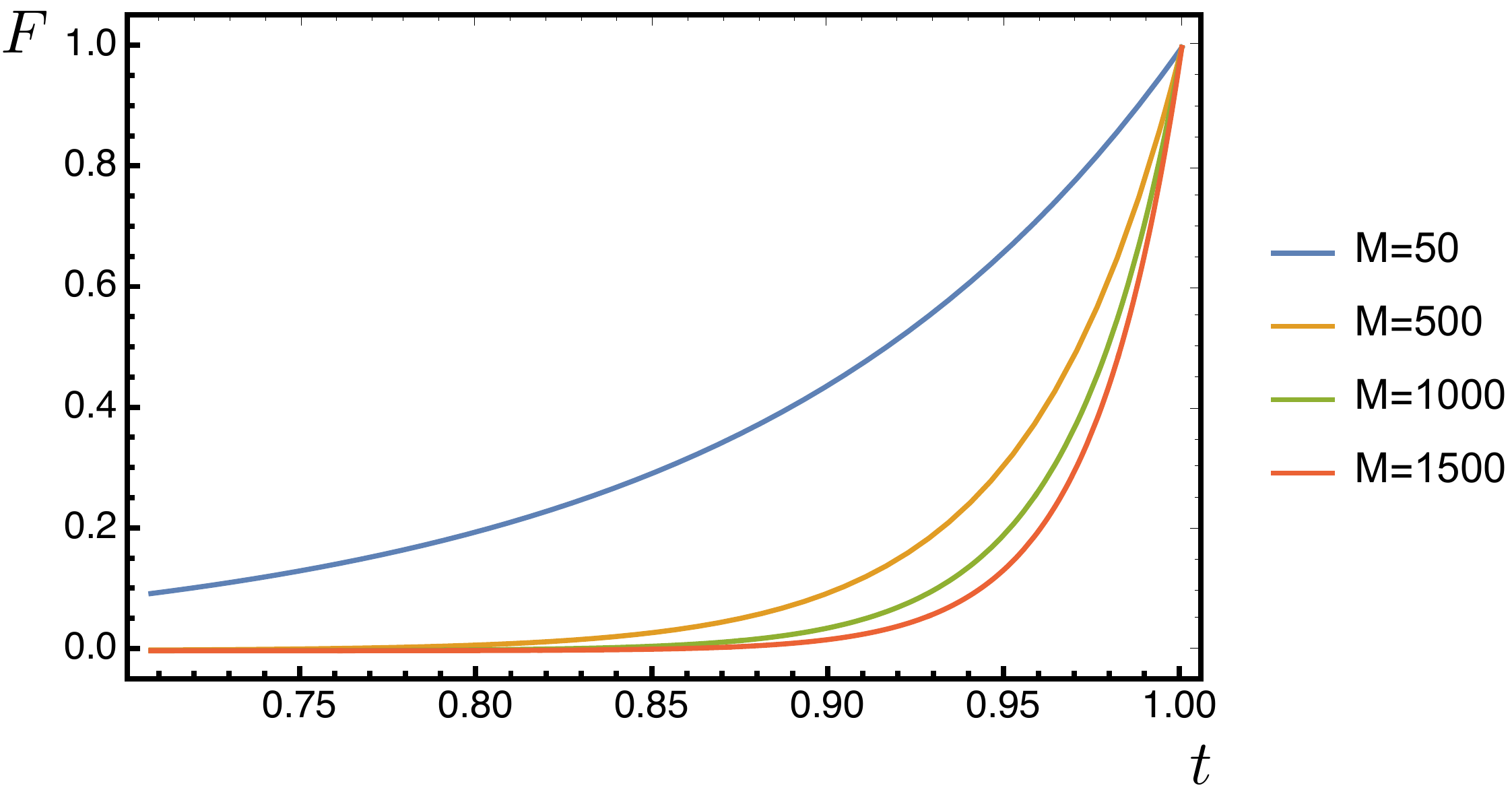}
	\caption{$F$ is the fidelity~(\ref{eq:qfunifloss}) between the ideal state, $\rho_{AB|\bm{U}}$, which is the (pure) state shared by Alice and Bob after Bob inputs his half of the $M$ two-mode squeezed-vacuum states into the ideal {\LON} described by transfer matrix $\bm{U}$, and the corresponding lossy state, $\rho_{AB|\bm{L}}$, for the case that the {\LON} has uniform loss, i.e., has transfer matrix $\bm{L}=t\bm{U}$, with $t$ being the uniform transmissivity.  The plots assume $\chi^2\!=1/\sqrt{M}$.  As shown, the fidelity drops exponentially as the transmissivity decreases.  For $15\%$ loss ($t^2=0.85$) the fidelity reduces to around $53\%$ for 50 modes, $16\%$ for 500 modes, $8\%$ for 1000 modes and below $5\%$ for 1500 modes.}
\label{fig:LBFwithLoss}
\end{figure}

We can now formulate the results of this subsection as the following condition: If after characterization of Bob's \LON, Alice finds that the right-hand side of Eq.~(\ref{eq:tvdbound}) is $\le\epsilon$, i.e.,
\begin{align}
\fe=\big[\qf\big]^2\ge1-\epsilon^2\,,
\end{align}
then the performance of Bob's \LON\ is acceptable.  This does not mean, however, as we have discussed, that the photocount samples are efficiently distinguishable from ones that can be generated by a classical algorithm.

As a specific example of our bounds, consider the very special case $\bm{L}=t\bm{U}$, where $t$ represents an overall, uniform
transmissivity through the \LON\ and ranges from $0$ to $1$, with 1 being no loss.  In this case the fidelity becomes
\begin{align}\label{eq:qfunifloss}
%\bc=
\qf
=\left(\frac{1-\chi^2}{1-\chi^2 t}\right)^M
\simeq e^{-M\chi^2(1- t)}e^{-M\chi^4(1- t^2)/2}\,.
\end{align}
Figure~\ref{fig:LBFwithLoss} plots how this fidelity scales with the transmissivity $t$, for different values of $M$.

\section{Discussion and conclusion}
\label{sec:discussion}

We present in this paper a protocol for characterizing randomized boson sampling that is implemented jointly by two parties, Alice and Bob, who share weakly squeezed, quantum-entangled, two-mode squeezed-vacuum states produced by a set of spontaneous parametric down-converters.  Alice instructs Bob to construct a linear-optical network into which he inputs his half of the two-mode squeezed-vacuum states.  Bob performs photon-counting measurements on the outputs of the \LON\ and reports the results to Alice. Alice performs either photon-counting measurements or heterodyne measurements on her half of the squeezed-vacuum states.  In the case of photocounting, it is reasonably likely that, given Alice's photocount record, Bob samples from a boson-sampling distribution with random single-photon inputs to the \LON.  In the case of heterodyne measurements, Alice can use the results, in combination with Bob's photocount record, to characterize {\it in situ\/} the transfer matrix of the \LON\ that Bob implemented, including losses at the input and through the \LON\ and sub-unit efficiency of the photodetectors.

Once Alice has characterized Bob's \LON, she can compare the RBS distribution for the ideal network she asked Bob to construct to the photocount distribution for the lossy network Bob managed to implement.  In developing measures for comparing these distributions, we find that the natural measure for comparing the ideal and lossy \LON s is the fidelity between the state shared by Alice and Bob after the ideal \LON\ and the shared state after Bob's lossy \LON; this fidelity can equivalently be thought of as the entanglement fidelity of a thermal state processed first through the lossy network and then through the inverse of the ideal network.  In our work, this natural measure is important because it provides an upper bound on the total variation distance between the ideal RBS distributions and the photocount distributions at the output of Bob's lossy \LON.  This measure has a simple expression in terms of the transfer matrices for the ideal and lossy \LON s; along with the related fidelity measures developed in App.~\ref{sec:fidelityboundsfixedN}, it might find application in other quantum protocols involving linear-optical networks.

Our characterization method is different from methods that require changing the measurement strategy at both ends in order to estimate a lower bound on the fidelity between the output state and the target state~\cite{Aolita2015} or to characterize the output state~\cite{Abrahao2018}.  Indeed, the key idea of our method is that by changing the measurement strategy from photocounting to heterodyne at Alice's end, without the need for any changes at Bob's end or Bob's even knowing what is going on, Alice converts RBS, a classically hard problem, to a problem that is classically efficiently simulable and, more importantly, enables complete, efficient characterization of Bob's \LON.  This strategy, changing measurements on an entangled system to convert a classically hard problem to one that is easy and also can be used to characterize the quantum device that does the hard problem, might be of interest for other quantum technologies.

We stress that not all kinds of imperfections at Bob's end can be dealt with by our characterization procedure.  In particular, dark counts in the photodetectors and mode-mismatched photons that, though lost from the correct mode, nonetheless make their way through the \LON\ to be counted by the photodetectors---these are somewhat like dark counts~\cite{RahimiKeshari2016}---are not described by a transfer matrix and thus cannot be handled by our current characterization procedure.  In addition, excess, presumably Gaussian noise introduced into Bob's network cannot be handled within our current formulation.  We plan to generalize our procedure to include such additional noise sources, both to certify whether such additional noise is present and, if it is, to find procedures to characterize both the noise and Bob's \LON.  In the case of mode-mismatched photons, however, no generalization of the transfer-matrix description to include mode mismatching is currently fully worked out.  Note also that the characterization runs, in practice, can introduce another mode-matching issue, that of matching the modes of local oscillators Alice uses for heterodyne detection to the mode shape of the modes Alice receives from the SPDC sources.

The RBS runs do not require the quantum correlations of the two-mode squeezed-vacuum state $\ket{\Psi_{AB}}$ of Eq.~(\ref{eq:2MSV}).  Indeed, as pointed out in~\cite{Shahandeh}, the RBS runs work in exactly the same way if one uses the state obtained by decohering $\ket{\Psi_{AB}}$ in the number basis, a so-called classical-classical state,
\begin{align}\label{eq:rhoCC}
\rho_{\textrm{CC}}
=(1-\chi^2)^M\sum_{\bm{n}_A}\chi^{2|\bm{n}_A|}\ketbra{\bm{n}_A}\otimes\ketbra{\bm{n}_B=\bm{n}_A}\,,
\end{align}
which has perfect, but purely classical photon-number correlations between the $A$ modes and the $B$ modes.  This state can be obtained from $\ket{\Psi_{AB}}$ by applying random phase operations to Alice's modes (or equivalently to Bob's modes before the LON); since photon-counting measurements are phase insensitive, this state leads to the boson-sampling distribution~(\ref{eq:lossyconditionalPrBS}), just as for the two-mode squeezed-vacuum state.

What the quantum entanglement of $\ket{\Psi_{AB}}$ enables is not the RBS runs, but rather the characterization runs.  A heterodyne outcome $\bm{\alpha}$ on the $A$ modes of $\ket{\Psi_{AB}}$ prepares the input to Bob's \LON\ in the coherent state $\ket{\bm{\chi\bm{\alpha}^*}}$ of Eq.~(\ref{eq:alphaonPsi}).  It is known that appropriately chosen coherent-state inputs to a \LON, with photocounting at the output, can be used to characterize the \LON~\cite{Rahimi-Keshari2013}; the ability of our protocol to prepare essentially random coherent-state inputs to Bob's \LON, an ability that comes from the quantum entanglement in the two-mode squeezed-vacuum state, is what allows our characterization protocol to work.  In contrast, given the classically correlated state~(\ref{eq:rhoCC}), the state input to the $B$ modes after a heterodyne measurement, is proportional to
\begin{align}\label{eq:rhoCChet}
\frac{1}{\pi^M}\bra{\bm{\alpha}}\rho_{\textrm{CC}}\ket{\bm{\alpha}}
=(1-\chi^2)^M\sum_{\bm{n}_B}\chi^{2|\bm{n}_B|}|\braket{\bm{\alpha}|\bm{n}_B}|^2\ketbra{\bm{n}_B}
=P_{\text{C}}(\bm{\alpha})\rho_{\text{CC}|\bm{\alpha}}\,.
\end{align}
Here $P_{\text{C}}(\bm{\alpha})$, the probability for heterodyne outcome $\bm{\alpha}$, is the same as for $\ket{\Psi_{AB}}$ and thus given by Eq.~(\ref{eq:probCalpha}), and the normalized state input to the \LON,
\begin{align}\label{eq:rhoCCalpha}
\rho_{\text{CC}|\bm{\alpha}}=\sum_{\bm{n}_B}\big\vert\braket{\bm{n}_B|\chi\bm{\alpha}^*}\big\vert^2\ketbra{\bm{n}_B}
=\int\frac{d\bm{\Phi}}{(2\pi)^M}\,\big\vert\chi(\bm{\alpha}e^{i\bm{\Phi}})^*\big\rangle\big\langle\chi(\bm{\alpha}e^{i\bm{\Phi}})^*\big\vert
\end{align}
is the coherent state $\ket{\chi\bm{\alpha}^*}$ phase-randomized and thus diagonal in the Fock basis.  The final form writes $\rho_{\text{CC}|\bm{\alpha}}$ in terms of a coherent-state expansion that is an explicit randomization of the phases, with $d\bm{\Phi}=d\phi_1\cdots d\phi_M$ denoting integration over the randomizing phases and $e^{i\bm{\Phi}}=\text{diag}(e^{i\phi_1},\ldots,e^{i\phi_M})$ being the diagonal unitary matrix formed from the randomizing phases.

The phase randomization means that $\rho_{\text{CC}}$ does not provide access to all the interference effects available when using two-mode squeezed vacuum as input.  In particular, it is obvious that the phase randomization wipes out the first-order coherence, i.e., the interference between complex amplitudes of Alice's modes, which is the basis for the characterization procedure of Sec.~\ref{sec:characterizationdetail} and is expressed in the second moments of Eq.~(\ref{eq:alphajki}).  Nonetheless, as we show in App.~\ref{sec:CCcharacterization}, the CC state does provide, in principle, nearly completely information about the transfer matrix---what is missing is that the heterodyne statistics cannot distinguish $\bm{L}$ from its complex conjugate $\bm{L}^*$, i.e., cannot remove an overall phase-reversal ambiguity in $\bm{L}$---but this information, because it is contained in coherences higher than first-order, is difficult to extract from the heterodyne data.

Notice now that even with the state $\rho_{\textrm{CC}}$, one can use Bob's unconditional photostatistics to recover the lossy part, $\sqrt{\bm{L}^\dagger\bm{L}}$, of the transfer matrix~(\ref{eq:pdL}), just as we discussed in Sec.~\ref{sec:Bobstats}.  We stress that the ability to get at this output loss matrix from Bob's unconditional photostatistics is a capacity made available by {\it randomized\/} boson sampling; this facility is not available in the original boson sampling, which injects single photons into a fixed set of input ports of the \LON.

It has been suggested~\cite{Shahandeh} that since there is evidently something nonclassical going on in the RBS runs, something that is not captured by thinking in terms of the classical correlations between photon-number eigenstates in the CC~state~(\ref{eq:rhoCC}), this nonclassicality might be that of quantum optics, arising from the nonclassical phase-space description of the state $\rho_{\textrm{CC}}$, i.e., that it has a negative Glauber-Sudarshan $P$-function~\cite{Agudelo2013}.  Yet the photon-number correlations in $\rho_{\textrm{CC}}$ are just a photon-mode way of describing the correlations with any classical device at Alice's end that selects number-state inputs to Bob's \LON\ from a thermal probability distribution; such a classical device, e.g., flipping coins to select the inputs to Bob's \LON\ or using random-number generation on a classical computer, generally has no phase-space description.  In the case of the CC~state, the results of the coin flipping are stored in number states at Alice's end, which are orthogonal and thus distinguishable; this distinguishability, automatic if the coin flipping is regarded as a classical device, is what makes the correlations completely classical, not having even the quantum correlations of quantum discord~\cite{Zurek2000,Ollivier2001,Lang2011}.\footnote{Discordant states can be prepared by coin flipping, but if the discord is to be nonzero, the results of the coin flipping are stored in states at Alice's end that are not distinguishable, unlike the situation for the CC state.  This means that the results of the coin flipping, though they can be read off the coin, cannot be determined from the quantum system at Alice's end; more generally, as the definition of quantum discord makes clear, not all the information about the correlations in a discordant state can be determined from local measurements.}

The nonclassicality of RBS lies in what Aaronson and Arkhipov discovered, that the propagation of single photons through a \LON\ produces a photon-number distribution that cannot be simulated efficiently classically.  There is no need for some other, additional nonclassicality to account for how the inputs to Bob's \LON\ are populated.  Indeed, what we contend is that in the version of RBS that uses two-mode squeezed-vacuum states, the genuine entanglement of these states is incidental to the nonclassicality of \hbox{RBS}, but instead plays the crucial role of enabling the ability to characterize the \LON\ from first-order coherence when Alice switches from photocounting to heterodyne measurement.

We close with a story.  Alice and Bob, having successfully run their randomized boson-sampling experiment, are in Stockholm to pick up the Nobel Prize in Physics for a demonstration of quantum supremacy.  As the King of Sweden approaches to present the Nobel medals, Alice rummages through her backpack to find the famous data to present to the King.  Suddenly, she cries out in alarm, ``I have the data from the characterization runs, so I know Bob implemented the \LON\ I requested to good accuracy, but I have lost the records of my photocounts in the RBS runs.  Without those records, Bob was sampling from a thermal distribution processed through his network.''  The startled King turns to the secretary of the Swedish Royal Academy of Sciences, who gives a thumbs down, remarking that he samples from such a distribution every time he goes outside on a sunny day.  As Bob and Alice beat a hasty retreat, the Nobel ceremony descends into chaos.

Is there a moral here?  There is a clich\'e: always preserve your data.  There is perhaps some discomfort with experiments that rely on post-selection.  But beyond clich\'e and post-selected queasiness might be something more.  The probability for a particular input $\ket{\bm{n}_A}$ into Bob's \LON\ is
\begin{align}
P_{\textrm{Q}}(\bm{n}_A)=(1-\chi^2)^M\chi^{2N}\simeq\chi^{2N}e^{-M\chi^2}e^{-M\chi^4/2}\sim\frac{1}{\sqrt e}(\sqrt{M}e)^{-\sqrt M}\,.
\end{align}
The third form assumes weak squeezing, $\chi^2\alt1/\sqrt M$.  The final form makes the standard choice, $\chi^2=1/\sqrt M$, and for that choice uses the typical number of photons, $N=\sqrt M$.  No matter how you slice it, the probability for any particular input and thus for any particular boson-sampling distribution at the output of the \LON\ decreases exponentially with problem size.  To sample twice from the same boson-sampling distribution thus takes an exponential number of runs, so verification that any particular sampling distribution is hard, no matter how efficient the verification test, takes an exponential number of runs.  Characterization of the sort proposed in this paper, which can provide confidence that Bob is sampling from something close to the ideal distribution, is almost certainly the best one can hope to do.  Randomized boson sampling that uses two-mode squeezed-vacuum states makes exponentially harder the already problematic task of verifying that one is doing a hard sampling problem, but provides the capability for {\it in situ\/} characterization of a lossy linear-optical network.  Now there is a moral.

\acknowledgments
This work was supported in part by NSF Grant No.~PHY-1630114.  We acknowledge useful discussions with R.~Alexander, C.~Jackson, E.~Kashefi, A.~Leverrier, D.~Markham, T.~J. Osborne, C.~Silberhorne, R.~Werner, and A.~Winter.  S.R.-K. acknowledges financial support from the German Academic Exchange Service (DAAD).

%%%%%%%%%%%%%%%%%%%%%%%%%%%%%%%%%%%%%%%%%%%%%%%%%%%%%%%%%%%%%%%%%%%%%%%%%%
%%%%%%%%%%%%%%%%%%%%%%%%%%%%%%%%%%%%%%%%%%%%%%%%%%%%%%%%%%%%%%%%%%%%%%%%%%
%%%%%%%%%%%%%%%%%%%%%%%%%%%%%%%%%%%%%%%%%%%%%%%%%%%%%%%%%%%%%%%%%%%%%%%%%%
%%%%%%%%%%%%%%%%%%%%%%%%%%%%%%%%%%%%%%%%%%%%%%%%%%%%%%%%%%%%%%%%%%%%%%%%%%
%%%%%%%%%%%%%%%%%%%%%%%%%%%%%%%%%%%%%%%%%%%%%%%%%%%%%%%%%%%%%%%%%%%%%%%%%%
%%%%%%%%%%%%%%%%%%%%%%%%%%%%%%%%%%%%%%%%%%%%%%%%%%%%%%%%%%%%%%%%%%%%%%%%%%

\appendix

\section{Lossy networks and fidelity bounds}
\label{sec:fidelitylowerbound}

In this Appendix we derive the fidelity bound and the explicit form of the fidelity given in Eq.~(\ref{eq:fidelitybound}).  In App.~\ref{sec:fidelityboundsfixedN} we proceed to related fidelity bounds that, instead of averaging over all input photocount records $\bm{n}_A$, average only over photon-number sectors with a fixed total number of photons.

\subsection{Lossy networks using auxiliary loss modes}
\label{sec:auxiliarymodes}

To make progress, we need to do some preliminary work by developing the description of the lossy network as part of a larger, lossless network that has $\tilde M$ auxiliary modes; the auxiliary modes are initialized in vacuum and receive the photons lost from Bob's original network.  This larger network is characterized by a unitary operator $\tilde{\mathcal{U}}$ and a corresponding unitary transfer matrix
\begin{align}\label{eq:tildeU}
\tilde{\bm{U}}=
\begin{pmatrix}
\bm{L}&\bm{R}\\\bm{S}&\tilde{\bm{L}}
\end{pmatrix}\,,
\end{align}
which describes, as in Eq.~(\ref{eq:Ubtransform}), how the larger network transforms the creation operators.  In Eq.~(\ref{eq:tildeU}), $\tilde{\bm{U}}$ is divided up according to the original $B$ modes, $M$ in number, and the auxiliary $\tilde B$ modes, $\tilde M$ in number.  The four submatrices are interrelated in various ways by the fact that $\bm{U}$ is unitary, but the only one of these constraints we need is that
\begin{align}\label{eq:LRconstraint}
\bm{L}\bm{L}^\dagger+\bm{R}\bm{R}^\dagger=\bm{I}\,.
\end{align}
It is useful below to consider the polar decomposition~(\ref{eq:pdL}) of the lossy transfer matrix.  It is also useful to examine the $M\times\tilde M$ matrix $\bm{R}$.  Without loss of generality, we can let the number of auxiliary $\tilde B$ modes equal the number of $B$ modes, i.e., $\tilde M=M$.  Then we have a polar decomposition,
\begin{align}\label{eq:pdR}
\bm{R}=\sqrt{\bm{R}\bm{R}^\dagger}\,\bm{O}=\sqrt{\bm{I}-\bm{L}\bm{L}^\dagger}\,\bm{O}\,,
\end{align}
in which we can always choose the unitary matrix $\bm{O}$, by redefining the auxiliary $\tilde B$ modes, to be the identity, in which case
$\bm{R}=\sqrt{\bm{I}-\bm{L}\bm{L}^\dagger}$.

The transfer matrix $\tilde{\bm{U}}$ describes how the larger network takes coherent states to coherent states, i.e.,
\begin{align}
\tilde{\mathcal{U}}\ket{\bm{\beta},\tilde{\bm{\beta}}}
=\big\vert\begin{pmatrix}\bm{\beta}&\tilde{\bm{\beta}}\end{pmatrix}\tilde{\bm{U}}\big\rangle
=\ket{\bm{\beta}\bm{L}+\tilde{\bm{\beta}}\bm{S},\bm{\beta}\bm{R}+\tilde{\bm{\beta}}\tilde{\bm{L}}}\,.
\end{align}
In the case of interest, the auxiliary modes begin in the vacuum state, symbolized by $\tilde{\bm{\beta}}=\tilde{\bm{0}}$, which gives
\begin{align}
\tilde{\mathcal{U}}\ket{\bm{\beta},\tilde{\bm{0}}}=\ket{\bm{\beta}\bm{L},\bm{\beta}\bm{R}}\,.
\end{align}
Equation~(\ref{eq:ELbeta}), for the action of the quantum operation of Bob's network on coherent states, follows immediately:
\begin{align}
\label{eq:ELbeta2}
\mathcal{E}_{\bm{L}}\big(\ketbra{\bm{\beta}}\big)
=\trace_{\tilde B}\big[\tilde{\mathcal{U}}\ketbra{\bm{\beta},\tilde{\bm{0}}}\tilde{\mathcal{U}}^\dagger\big]
=\ketbra{\bm{\beta L}}\,.
\end{align}
We can write the quantum operation $\mathcal{E}_{\bm{L}}$ in a way better suited to Eq.~(\ref{eq:bcBS}) by taking the trace over the auxiliary modes in the number basis,
\begin{align}\label{eq:ELKraus}
\mathcal{E}_{\bm{L}}(\rho_B)
=\trace_{\tilde B}\big[\tilde{\mathcal{U}}\rho_{B}\otimes\ketbra{\tilde{\bm{0}}}\tilde{\mathcal{U}}^\dagger\big]
=\sum_{\bm{n}_{\tilde B}}
\bra{\bm{n}_{\tilde B}}\tilde{\mathcal{U}}\ket{\tilde{\bm{0}}}
\rho_B
\bra{\tilde{\bm{0}}}\tilde{\mathcal{U}}^\dagger\ket{\tilde{\bm{n}}_{\tilde B}}
=\sum_{\bm{n}_{\tilde B}}\mathcal{K}_{\bm{n}_{\tilde B}}\rho_B \mathcal{K}_{\bm{n}_{\tilde B}}^\dagger\,,
\end{align}
where in the final form, we recognize Kraus operators that act on the $B$ modes,
\begin{align}
\mathcal{K}_{\bm{n}_{\tilde B}}=\bra{\bm{n}_{\tilde B}}\tilde{\mathcal{U}}\ket{\tilde{\bm{0}}}\,.
\end{align}
To get back to how $\mathcal{E}_{\bm{L}}$ acts on coherent states, we can use that the Kraus operators act on coherent states according to
\begin{align}\label{eq:Kncoh}
\mathcal{K}_{\bm{n}_{\tilde B}}\ket{\bm{\beta}}
=\bra{\bm{n}_{\tilde B}}\tilde{\mathcal{U}}\ket{\bm{\beta},\tilde{\bm{0}}}
=\ket{\bm{\beta}\bm{L}}\braket{\bm{n}_{\tilde B}|\bm{\beta}\bm{R}}\,,
\end{align}
which is just what Bob's \LON\ does, except that each Kraus operator includes the square root of the probability, $\vert\braket{\bm{n}_{\tilde B}|\bm{\beta}\bm{R}}\vert^2$, for the photon record $\bm{n}_{\tilde B}$ to be left in the auxiliary modes.  There is no linear operator that takes $\ket{\bm{\beta}}$ to $\ket{\bm{\beta}\bm{L}}$ when $\bm{L}$ is not unitary; the Kraus operators do this job by sub-normalizing the output coherent state in a particular way.

From here on we omit the reference to the auxiliary modes in the subscript on the Kraus operators, writing the subscript as $\bm{n}$, and simply remembering that this is a record of counts left in the auxiliary modes.

What we need shortly is just one special case of Eq.~(\ref{eq:Kncoh}), for the Kraus operator
\begin{align}\label{eq:K0}
\mathcal{K}_{\bm{0}}=\bra{\tilde{\bm{0}}}\tilde{\mathcal{U}}\ket{\tilde{\bm{0}}}\,,
\end{align}
which corresponds to no photons lost into the auxiliary modes; for this special case, we have
\begin{align}\label{eq:K0beta}
\mathcal{K}_{\bm{0}}\ket{\bm{\beta}}
=\ket{\bm{\beta}\bm{L}}\braket{\tilde{\bm{0}}|\bm{\beta}\bm{R}}
=\ket{\bm{\beta}\bm{L}}e^{-\bm{\beta}\bm{R}\bm{R}^\dagger\bm{\beta}^\dagger/2}
=\ket{\bm{\beta}\bm{L}}e^{-\bm{\beta}(\bm{I}-\bm{L}\bm{L}^\dagger)\bm{\beta}^\dagger/2}\,,
\end{align}
where we use the unitarity constraint~(\ref{eq:LRconstraint}) to write the final expression in terms of the lossy transfer matrix $\bm{L}$.

We can manipulate Eq.~(\ref{eq:Kncoh}) so that it gives an explicit expression for all the Kraus operators in terms of $\mathcal{K}_{\bm{0}}$,
\begin{align}\label{eq:Kncoh2}
\begin{split}
\mathcal{K}_{\bm{n}}\ket{\bm{\beta}}
&=\ket{\bm{\beta}\bm{L}}e^{-\bm{\beta}\bm{R}\bm{R}^\dagger\bm{\beta}^\dagger/2}
\prod_{j=1}^M\frac{\big[(\bm{\beta}\bm{R})_j\big]^{n_j}}{n_j!}\\
&=\mathcal{K}_{\bm{0}}\ket{\bm{\beta}}
\prod_{j=1}^M\frac{\big[(\bm{\beta}\bm{R})_j\big]^{n_j}}{n_j!},\\
&=\mathcal{K}_{\bm{0}}\prod_{j=1}^M\frac{\big[(\bm{b}\bm{R})_j\big]^{n_j}}{n_j!}\ket{\bm{\beta}}\,,
\end{split}
\end{align}
where $\bm{b}$ is the row vector~(\ref{eq:rowb}) of annihilation operators.  We can read off an explicit form for $\mathcal{K}_{\bm{n}}$, which we display below in Eq.~(\ref{eq:Kn}).

It is both useful and instructive to find an explicit expression for $\mathcal{K}_{\bm{0}}$.  Even though we can do without this explicit form in what follows, it elucidates the nature of the lossy \LON.  Let $\mathcal{V}$ be the unitary operator that implements the lossless transfer matrix~$\bm{V}$,~i.e.,
\begin{align}\label{eq:Vaction}
\mathcal{V}\ket{\bm{\beta}}=\ket{\bm{\beta}\bm{V}}\,,
\end{align}
for any coherent state $\ket{\bm{\beta}}$.  We recall the formula
\begin{align}\label{eq:Laction}
e^{-\bm{\beta}(\bm{I}-\bm{L}\bm{L}^\dagger)\bm{\beta}^\dagger/2}\big\vert\bm{\beta}\sqrt{\bm{L}\bm{L}^\dagger}\big\rangle
=\mathcal{L}\ket{\bm{\beta}}\,,
\end{align}
where
\begin{align}
\mathcal{L}=\exp\!\bigg(\sum_{j,k}H_{jk}b_k^\dagger b_j\bigg)
\end{align}
is a Hermitian, photon-number-conserving operator on the $B$ modes, with
\begin{align}
\bm{H}=\ln\big(\sqrt{\bm{L}\bm{L}^\dagger}\big)\,.
\end{align}
Combining Eqs.~(\ref{eq:Vaction}) and~(\ref{eq:Laction}) into Eq.~(\ref{eq:K0beta}) gives
\begin{align}
\mathcal{V}\mathcal{L}\ket{\bm{\beta}}=\ket{\bm{\beta}\bm{L}}e^{-\bm{\beta}(\bm{I}-\bm{L}\bm{L}^\dagger)\bm{\beta}^\dagger/2}
=\mathcal{K}_{\bm{0}}\ket{\bm{\beta}}\,.
\end{align}
which implies that
\begin{align}
\mathcal{K}_{\bm0}=\mathcal{V}\mathcal{L}.
\end{align}
The polar decomposition~(\ref{eq:pdL}) of $\bm{L}$ thus leads to a corresponding polar decomposition of $\mathcal{K}_{\bm{0}}$: $\mathcal{V}$ is the unitary operator for the lossless network $\bm{V}$;
\begin{align}
\mathcal{L}=\sqrt{\mathcal{K}_{\bm0}^\dagger\mathcal{K}_{\bm0}}
\end{align}
comes from the remaining part of the polar decomposition, $\sqrt{\bm{L}\bm{L}^\dagger}$, which describes the losses.

Formula~(\ref{eq:Laction}) is the multimode generalization of the single-mode formula
\begin{align}
e^{-(1-\lambda^2)|\alpha|^2/2}\ket{\lambda\alpha}=\lambda^{a^\dagger a}\ket\alpha\,,
\end{align}
which proves useful whenever one considers a quantum process that take coherent states to coherent states.  It is often used to provide the single-mode Kraus-operator description of losses~\cite{Wiseman}, for which the present description is the multimode generalization, and it has been used productively in analyzing nondeterministic linear amplifiers~\cite{Pandey}, where $\lambda^{a^\dagger a}$ emerges, just as here, in a Kraus operator, one that describes, when $\lambda>1$, probabilistic amplification of an initial coherent state to a coherent state of higher amplitude.

Returning now to the Kraus operators, we extract the explicit form of $\mathcal{K}_{\bm{n}}$ from Eq.~(\ref{eq:Kncoh2}),
\begin{align}\label{eq:Kn}
\mathcal{K}_{\bm{n}}
=\mathcal{K}_{\bm{0}}\prod_{j=1}^M\frac{\big[\big(\bm{b}\sqrt{\bm{I}-\bm{L}\bm{L}^\dagger}\,\big)_j\big]^{n_j}}{n_j!}
=\prod_{j=1}^M\frac{\big[\big(\bm{b}\bm{L}^{-1}\sqrt{\bm{I}-\bm{L}\bm{L}^\dagger}\big)_j\big]^{n_j}}{n_j!}\mathcal{K}_{\bm{0}}\,.
\end{align}
Here the latter form uses
\begin{align}
\mathcal{K}_{\bm{0}}^{-1}\bm{b}\mathcal{K}_{\bm{0}}=\bm{b}\bm{L},
\end{align}
which follows immediately from the action~(\ref{eq:K0}) of $\mathcal{K}_{\bm{0}}$ on coherent states (likewise, we have $\mathcal{V}^\dagger\bm{b}\mathcal{V}=\bm{b}\bm{V}$ and $\mathcal{L}^{-1}\bm{b}\mathcal{L}=\bm{b}e^{\bm{H}}=\bm{b}\sqrt{\bm{L}\bm{L}^\dagger}$).

\subsection{Fidelity bound~(\ref{eq:fidelitybound})}
\label{sec:thefidelitybound}

We turn now to the fidelity bound~(\ref{eq:fidelitybound}) and start with the fidelity~(\ref{eq:bcBS}) between the ideal and lossy RBS distributions,
\begin{align}\label{eq:bcBS2}
\begin{split}
\bcBS&=\sum_{\bm{n}_B}\sqrt{P_{\textrm{BS}}(\bm{n}_B|\bm{n}_A,\bm{U})P_{\textrm{BS}}(\bm{n}_B|\bm{n}_A,\bm{L})}\\
&=\hspace{-4pt}\sum_{{\scriptstyle{\bm{n}_B}\atop\scriptstyle{|\bm{n}_B|=|\bm{n}_A|}}}\hspace{-4pt}
|\bra{\bm{n}_A}\mathcal{U}^\dagger\ket{\bm{n}_B}|\,\sqrt{\bra{\bm{n}_B} \mathcal{E}_{L}\left(\ketbra{\bm{n}_A}\right) \ket{\bm{n}_B}}\,.
\end{split}
\end{align}
The factor $\big\vert\bra{\bm{n}_B} \mathcal{U} \ket{\bm{n}_A}\big\vert$ is zero unless $|\bm{n}_B|=|\bm{n}_A|$, so it gives the indicated restriction on the sum.  We remind the reader that although $\bm{n}_A$ is the record of photocounts at Alice's end, it is better thought of here as specifying the state input to Bob's \LON.  Indeed, the analysis here is carried out within the state space of the $B$ modes and the auxiliary $\tilde B$ modes; the only time Alice's end comes into play is in relating the fidelity bound to the quantum fidelity of Eq.~(\ref{eq:fidelitybound}).

We now substitute the Kraus-operator expansion~(\ref{eq:ELKraus}) of $\mathcal{E}_{\bm{L}}$ into the fidelity~(\ref{eq:bcBS2}), noting that the restriction $|\bm{n}_B|=|\bm{n}_A|$ means that we only need to keep the single Kraus term that corresponds to no photons leaking into the auxiliary modes, thus obtaining
\begin{align}\label{eq:bcBS3}
\bcBS
=\hspace{-4pt}\sum_{{\scriptstyle{\bm{n}_B}\atop\scriptstyle{|\bm{n}_B|=|\bm{n}_A|}}}\hspace{-4pt}
\big\vert\bra{\bm{n}_A}\mathcal{U}^\dagger\ket{\bm{n}_B}\bra{\bm{n}_B}\mathcal{K}_{\bm{0}}\ket{\bm{n}_A}\big\vert\,,
\end{align}
where
\begin{align}
\bra{\bm{n}_B}\mathcal{K}_{\bm{0}}\ket{\bm{n}_A}=\bra{\bm{n}_B,\tilde{\bm{0}}}\tilde{\mathcal{U}}\ket{\bm{n}_A,\tilde{\bm{0}}}\,.
\end{align}
From here on, we omit the restriction $|\bm{n}_B|=|\bm{n}_A|$, it being enforced by both factors in the sum.  Notice that if Bob's network is the desired, ideal one, we can choose $\tilde{\mathcal{U}}=\mathcal{U}\otimes\mathcal{I}$, which means that $\mathcal{K}_{\bm{0}}=\mathcal{U}$ is the only nonzero Kraus operator, and the classical fidelity~(\ref{eq:bcBS3}) is 1, as it should be, since $\bm{L}=\bm{U}$.

Equation~(\ref{eq:bcBS3}) is an equality, not a bound.  We now take the step away from equality to a bound by pulling the absolute value outside the sum:
\begin{align}\label{eq:BSbound}
\begin{split}
\bcBS
&\ge\bigg\vert\sum_{\bm{n}_B}
\bra{\bm{n}_A}\mathcal{U}^\dagger\ket{\bm{n}_B}\bra{\bm{n}_B}\mathcal{K}_{\bm0}\ket{\bm{n}_A}\bigg\vert\\
&=\Big\vert\bra{\bm{n}_A}\mathcal{U}^\dagger\,\mathcal{K}_{\bm0}\ket{\bm{n}_A}\Big\vert\\
&=F\Big(\mathcal{U}\ket{\bm{n}_A},\mathcal{E}_{\bm{L}}\big(\ketbra{\bm{n}_A}\big)\Big)\,.
\end{split}\end{align}
The condition for equality here is that all the terms in the sum have the same phase.

The bound is now in a very neat form---it is the fidelity between the states $\mathcal{U}\ket{\bm{n}_A}$ and $\mathcal{E}_{\bm{L}}\big(\ketbra{\bm{n}_A}\big)$ or, equivalently, the overlap of the states $\mathcal{U}\ket{\bm{n}_A}$ and $\mathcal{K}_{\bm{0}}\ket{\bm{n}_A}=\mathcal{V}\mathcal{L}\ket{\bm{n}_A}$---but to get a bound involving the transfer matrices $\bm{U}$ and $\bm{L}$ requires further massaging.  The first step is to average over $P_{\textrm{Q}}(\bm{n}_A)$, as in Eq.~(\ref{eq:bc}) (in App.~\ref{sec:fidelityboundsfixedN} we formulate bounds based on averaging over fixed-photon-number sectors).  This gives us
\begin{align}\label{eq:fidelitybound2}
\begin{split}
\bc&=\sum_{\bm{n}_A}P_{\textrm{Q}}(\bm{n}_A)\bcBS\\
&\ge\sum_{\bm{n}_A}\big\vert P_{\textrm{Q}}(\bm{n}_A)\bra{\bm{n}_A}\mathcal{U}^\dagger\,\mathcal{K}_{\bm0}\ket{\bm{n}_A}\big\vert\\
&\ge\bigg\vert\sum_{\bm{n}_A}P_{\textrm{Q}}(\bm{n}_A)\bra{\bm{n}_A}\mathcal{U}^\dagger\,\mathcal{K}_{\bm0}\ket{\bm{n}_A}\bigg\vert\\
&=\bigg\vert\,\trace\bigg[\bigg(\sum_{\bm{n}_A}(1-\chi^2)^M \chi^{2|\bm{n}_A|}\ketbra{\bm{n}_A}\bigg)\mathcal{U}^\dagger\,\mathcal{K}_{\bm0}\bigg]\bigg\vert\\
&=\Big\vert\,\trace\big[\rho_{\textrm{th},B}\,\mathcal{U}^\dagger\,\mathcal{K}_{\bm0}\big]\Big\vert\,,
\end{split}
\end{align}
where we use that the state in large parentheses in the fourth line is the thermal state of Eq.~(\ref{eq:rhoth}).  The condition for equality in the third line is that all the terms $\bra{\bm{n}_A}\mathcal{U}^\dagger\,\mathcal{K}_{\bm0}\ket{\bm{n}_A}$ have the same phase.  A little reflection shows that the overall condition for equality is an extension of that expressed above, that the terms $\bra{\bm{n}_A}\mathcal{U}^\dagger\ket{\bm{n}_B}\bra{\bm{n}_B}\mathcal{K}_{\bm0}\ket{\bm{n}_A}$ have the same phase for all input and output photocount records.  The bound is now in terms of the operator overlap of $\mathcal{U}$ and $\mathcal{K}_{\bm{0}}=\mathcal{V}\mathcal{L}$, as measured in the thermal state.

We now show in a rush that the bound~(\ref{eq:fidelitybound2}) is the quantum fidelity of Eq.~(\ref{eq:fidelitybound}),
\begin{align}\label{eq:boundtoqf}
\begin{split}
\qf
&=\big\langle\Psi_{AB|\bm{U}}\big|\rho_{AB|\bm{L}}\big|\Psi_{AB|\bm{U}}\big\rangle^{1/2}\\
&=\Big\langle\Psi_{AB}\Big|
\mathcal{I}_A\otimes\mathcal{U}^\dagger\mathcal{E}_{\bm L}\big(\ketbra{\Psi_{AB}}\big)\mathcal{I}_A\otimes\mathcal{U}
\Big|\Psi_{AB}\Big\rangle^{1/2}\\
&=\big\vert\bra{\Psi_{AB}}\mathcal{I}_A\otimes\mathcal{U}^\dagger\mathcal{K}_{\bm0}\ket{\Psi_{AB}}\big\vert\\
&=\big\vert\trace_{AB}\big[\mathcal{I}_A\otimes\mathcal{U}^\dagger\mathcal{K}_{\bm0}\ketbra{\Psi_{AB}}\big]\big\vert\\
&=\big\vert\trace_B\big[\mathcal{U}^\dagger\mathcal{K}_{\bm0}\rho_{\text{th},B}\big]\big\vert\,,
\end{split}
\end{align}
thus confirming, in this particular case, the general bound implicit in Eq.~(\ref{eq:qf}).

One final step is left, to write the fidelity bound in terms of the transfer matrices.  To do that, use the coherent-state expansion~(\ref{eq:Glauber-Sudarshan}) of the thermal state and Eq.~(\ref{eq:meannin}) to write
\begin{align}
\trace\big[\rho_{\textrm{th},B}\,\mathcal{U}^\dagger\,\mathcal{K}_{\bm0}\big]
=\left(\frac{1-\chi^2}{\chi^{2}}\right)^M\int\frac{d^{2M}\!\bm{\beta}}{\pi^M}\,
\exp\!\left(-\frac{1-\chi^2}{\chi^2}\bm{\beta}\bm{\beta}^\dagger\right)\bra{\bm{\beta}}\mathcal{U}^\dagger\,\mathcal{K}_{\bm0}\ket{\bm{\beta}}\,.
\end{align}
We are really cooking now because we can use Eqs.~(\ref{eq:Ubeta}) and~(\ref{eq:K0beta}) to convert the coherent-state matrix element to a form that involves the transfer matrices $\bm{U}$ and $\bm{L}$,
\begin{align}
\bra{\bm{\beta}}\mathcal{U}^\dagger\,\mathcal{K}_{\bm0}\ket{\bm{\beta}}
=e^{-\bm{\beta}\bm{\beta}^\dagger/2}e^{\bm{\beta}\bm{L}\bm{L}^\dagger\bm{\beta}^\dagger/2}\braket{\bm{\beta}\bm{U}|\bm{\beta}\bm{L}}
=e^{-\bm{\beta}(\bm{I}-\bm{L}\,\bm{U}^\dagger)\bm{\beta}^\dagger}\,.
\end{align}
What is left is a Gaussian integral,
\begin{align}\label{eq:gaussint}
\trace\big[\rho_{\textrm{th},B}\,\mathcal{U}^\dagger\,\mathcal{K}_{\bm0}\big]
=\left(\frac{1-\chi^2}{\chi^{2}}\right)^M
\int\frac{d^{2M}\!\bm{\beta}}{\pi^M}\,
\exp\!\left[-\bm{\beta}\left(\frac{1}{\chi^2}\bm{I}-\bm{L}\,\bm{U}^\dagger\right)\bm{\beta}^\dagger\right]
=\frac{(1-\chi^2)^M}{\det(\bm{I}-\chi^2\bm{L}\,\bm{U}^\dagger)}\,,
\end{align}
which puts the fidelity bound into its final form,
\begin{align}\label{eq:fidelitybound3}
\bc\ge\qf
=\big\vert\,\trace\big[\rho_{\textrm{th},B}\,\mathcal{U}^\dagger\,\mathcal{K}_{\bm0}\big]\big\vert
=\frac{(1-\chi^2)^M}{\big\vert\det(\bm{I}-\chi^2\bm{L}\,\bm{U}^\dagger)\big\vert}\,.
\end{align}
Notice that $\bm{L}\,\bm{U}^\dagger=\sqrt{\bm{L}\bm{L}^\dagger}\bm{V}\bm{U}^\dagger$, emphasizing again that the lossy network $\bm{L}$ is weighed against the ideal network in two ways based on its polar decomposition: first, on how well its associated unitary transfer matrix, $\bm{V}$, matches $\bm{U}$ and, second, on how close its nonunitary, lossy piece, $\sqrt{\bm{L}\bm{L}^\dagger}$, is to the identity matrix.

An important special case occurs when the lossless network that is associated with $\bm{L}$ is the same as the ideal network, i.e., $\bm{V}=\bm{U}$, which is equivalent to $\mathcal{V}=\mathcal{U}$.  In this case the quantum fidelity becomes
\begin{align}
\qf
=\big\vert\,\trace[\rho_{\textrm{th},B}\mathcal{L}]\big\vert
=\frac{(1-\chi^2)^M}{\det\!\big(\bm{I}-\chi^2\sqrt{\bm{L}\bm{L}^\dagger}\big)}
=\frac{(1-\chi^2)^M}{(1-\chi^2 t_1)\cdots(1-\chi^2 t_M)}\,,
\end{align}
where $t_i$ are the singular values of $\bm{L}$.  We note that when $\mathcal{V}=\mathcal{U}$, only the pure loss in Bob's \LON\ needs to be characterized, and that can be done using Bob's unconditional photostatistics, without any need for the heterodyne characterization runs.

An even more special case, which we consider in the main text (and plot in Fig.~\ref{fig:LBFwithLoss}), to get an idea of the scaling due to losses, occurs when $\bm{L}=t\,\bm{U}$.  Here $t$ is an overall, uniform transmissivity through the \LON.  In this case, $\bm{V}=\bm{U}$, $\sqrt{\bm{L}\bm{L}^\dagger}=t\bm{I}$, $\mathcal{V}=\mathcal{U}$, and
\begin{align}
\mathcal{L}=t^{\,\sum_j\!b_j^\dagger b_j}\,.
\end{align}
The fidelity bound is tight, since $\bra{\bm{n}_A}\mathcal{U}^\dagger\ket{\bm{n}_B} \bra{\bm{n}_B}\mathcal{K}_{\bm0}\ket{\bm{n}_A}=t^{|\bm{n}_A|}\vert\bra{\bm{n}_A}\ \mathcal{U}\ket{\bm{n}_B}\vert^2$ is always real and nonnegative.  Thus, in this very special case, we have
\begin{align}
\bc=\qf
=\Big\vert\,\trace\big[\rho_{\textrm{th},B}\,t^{\sum_jb_j^\dagger b_j}\big]\Big\vert
=\left(\frac{1-\chi^2}{1-\chi^2 t}\right)^M\,.
\end{align}

\subsection{Fidelity bounds in fixed photon-number sectors}
\label{sec:fidelityboundsfixedN}

In App.~\ref{sec:thefidelitybound} we formulate bounds on the classical fidelity between the joint distributions $P_{\text{Q}|\bm{U}}$ and $P_{\text{Q}|\bm{L}}$ of Eqs.~(\ref{eq:JointPrRBS}) and~(\ref{eq:lossyJointPrRBS}).  At Eq.~(\ref{eq:fidelitybound2}), this involved averaging the fidelity between the ideal and lossy boson-sampling distributions over all input photocount records $\bm{n}_A$.  In this section, we formulate fidelity bounds that involve averaging only over photon-number sectors with a fixed total number of photons $N=\vert\bm{n}_A\vert$.

To state these bounds compactly, we need to introduce additional states that correspond to knowing the total number of photons at the input to Bob's network.
We start by projecting the input two-mode squeezed state to the $N$-photon sector by using Eqs.~(\ref{eq:PiN}), (\ref{eq:trPiN}), and (\ref{eq:PQN}),
\begin{align}\label{eq:PsiABN}
\ket{\Psi_{AB|N}}
&=\frac{1}{\sqrt{P_Q(N)}}\Pi_N\otimes\Pi_N\ket{\Psi_{AB}}
=\frac{1}{\sqrt{G(N,M)}}\hspace{-5pt}\sum_{{\scriptstyle{\bm{n}_A}\atop\scriptstyle{|\bm{n}_A|=N}}}\hspace{-5pt}\ket{\bm{n}_A}\otimes\ket{\bm{n}_B=\bm{n}_A}\,,
\end{align}
and we let
\begin{align}
\rho_{AB|N}&=\ketbra{\Psi_{AB|N}}
\end{align}
denote the associated density operator.  The corresponding marginal state at the input to Bob's \LON\ is the microcanonical ensemble for a fixed number of photons,
\begin{align}
\rho_{B|N}=\trace_A[\rho_{AB|N}]
=\frac{1}{G(N,M)}\Pi_N
=\frac{1}{P_Q(N)}\Pi_N\rho_{\text{th},B}\Pi_N\,;
\end{align}
as indicated, this state is obtained by projecting the canonical-ensemble thermal state~(\ref{eq:marginalthermal}) into the $N$-photon sector.  The corresponding states after propagation through the ideal and lossy networks are
\begin{align}
\ket{\Psi_{AB|N,\bm{U}}}&=\mathcal{I}_A\otimes\mathcal{U}\ket{\Psi_{AB,N}}=\frac{1}{\sqrt{P_Q(N)}}\Pi_N\otimes\Pi_N\ket{\Psi_{AB|\bm{U}}}\,,\\
\label{eq:rhoABNU}
\rho_{AB|N,\bm{U}}&=\ketbra{\Psi_{AB|N,\bm{U}}}=\mathcal{I}_A\otimes\mathcal{E}_{\bm U}(\rho_{AB|N})\,,\\
\label{eq:rhoABNL}
\rho_{AB|N,\bm{L}}&=\mathcal{I}_A\otimes\mathcal{E}_{\bm L}(\rho_{AB|N})\,.\vphantom{\Big(}
\end{align}

In this section we are interested in the fidelity between the joint photocounting distributions, $P_{\text{Q}|N,\bm{U}}$ and $P_{\text{Q}|N,\bm{L}}$, constructed from the states~(\ref{eq:rhoABNU}) and~(\ref{eq:rhoABNL}),
\begin{align}
P_{\text{Q}}(\bm{n}_A,\bm{n}_B|N,\bm{U})
&=\vert\braket{\bm{n}_A,\bm{n}_B|\Psi_{AB|N,\bm{U}}}\vert^2
=P_{\text{Q}}(\bm{n}_A|N)\,P_{\text{BS}}(\bm{n}_B|\bm{n}_A,\bm U)\,,\\
P_{\text{Q}}(\bm{n}_A,\bm{n}_B|N,\bm{L})
&=\bra{\bm{n}_A,\bm{n}_B}\rho_{AB|N,\bm{L}}\ket{\bm{n}_A,\bm{n}_B}
=P_{\text{Q}}(\bm{n}_A|N)P_{\text{BS}}(\bm{n}_B|\bm{n}_A,\bm{L})\,,
\end{align}
where
\begin{align}
P_{\text{Q}}(\bm{n}_A|N)=\frac{\delta_{|\bm{n}_A|,N}}{G(N,M)}
\end{align}
is the probability for photocount record $\bm{n}_A$, given $N$ total photocounts, and $P_{\text{BS}}(\bm{n}_B|\bm{n}_A,\bm U)$ and $P_{\text{BS}}(\bm{n}_B|\bm{n}_A,\bm{L})$ are the ideal and lossy boson-sampling distributions of Eqs.~(\ref{eq:conditionalPrBS}) and~(\ref{eq:lossyconditionalPrBS}).

To get the desired fidelity bounds, we start with Eq.~(\ref{eq:BSbound}) and proceed through the same steps as in Eqs.~(\ref{eq:fidelitybound2}) and~(\ref{eq:boundtoqf}) to obtain
\begin{align}\label{eq:fidelityboundN}
\bcN=\sum_{\bm{n}_A}P_{\textrm{Q}}(\bm{n}_A|N)\bcBS
\ge\Big\vert\,\trace\big[\rho_{B|N}\,\mathcal{U}^\dagger\,\mathcal{K}_{\bm0}\big]\Big\vert
=\qfN\,.
\end{align}
The condition for equality is that the terms $\bra{\bm{n}_A}\mathcal{U}^\dagger\ket{\bm{n}_B}\bra{\bm{n}_B}\mathcal{K}_{\bm0}\ket{\bm{n}_A}$ have the same phase for all input and output photocount records with $|\bm{n}_A|=|\bm{n}_B|=N$.   It should now be clear that whereas the states introduced in Eqs.~(\ref{eq:PsiABN})--(\ref{eq:rhoABNL}) are not the states that are used in our RBS protocol, they are the states that arise naturally in considering the average fidelity between the ideal and lossy boson-sampling distributions when the average is restricted to a particular total photon number $N$.

The final step of writing the bound in terms of the transfer matrices follows from realizing that the marginal thermal state at the input to Bob's \LON,
\begin{align}
\rho_{\text{th},B}=\sum_{N=0}^\infty P_{\text{Q}}(N)\rho_{B|N}\,,
\end{align}
provides a generating function for the bounds~(\ref{eq:fidelityboundN}),
\begin{align}
\frac{1}{\det(\bm{I}-\chi^2\bm{L}\,\bm{U}^\dagger)}
=\frac{\trace\big[\rho_{\text{th},B}\,\mathcal{U}^\dagger\,\mathcal{K}_{\bm0}\big]}{(1-\chi^2)^M}
=\sum_{N=0}^\infty\chi^{2N}G(N,M)\trace\big[\rho_{B|N}\,\mathcal{U}^\dagger\,\mathcal{K}_{\bm0}\big]\,,
\end{align}
where the generating function is evaluated in terms of the transfer matrices using Eq.~(\ref{eq:gaussint}).  We end up with
\begin{align}\label{eq:gaussintN}
\trace\big[\rho_{B|N}\,\mathcal{U}^\dagger\,\mathcal{K}_{\bm0}\big]=
\frac{(M-1)!}{(N+M-1)!}\left.\frac{\partial^N}{\partial(\chi^2)^N}\frac{1}{\det(\bm{I}-\chi^2\bm{L}\,\bm{U}^\dagger)}\right|_{\chi^2=0}\,.
\end{align}

To illustrate what is going on, consider the very simple case of a single photon, $N=1$, input to Bob's \LON,
\begin{align}\label{eq:fidelitybound1}
F\big(P_{\textrm{Q}|1,\bm{U}}, P_{\textrm{Q}|1,\bm{L}}\big)
\ge\big\vert\,\trace\big[\rho_{B|1}\,\mathcal{U}^\dagger\,\mathcal{K}_{\bm0}\big]\big\vert
=\frac{1}{M}\big\vert\trace[\bm{L}\,\bm{U}^\dagger]\big\vert\,.
\end{align}
A single photon propagates through the network with quantum amplitudes that are the same as the coherent-state amplitudes, so it is not surprising that the fidelity bound reduces to an overlap of the transfer matrices that describe the propagation of coherent states.  When $\bm{V}=\bm{U}$, we have
\begin{align}
F\big(P_{\textrm{Q}|1,\bm{U}},P_{\textrm{Q}|1,\bm{L}}\big)
\ge\big\vert\,\trace\big[\rho_{B|1}\,\mathcal{U}^\dagger\,\mathcal{K}_{\bm0}\big]\big\vert
=\frac{1}{M}\sum_{i=1}^M t_i\,.
\end{align}
For the case of uniform loss, $\bm{L}=t\,\bm{U}$, the bound is saturated, and we have $F\big(P_{\textrm{Q}|1,\bm{U}}, P_{\textrm{Q}|1,\bm{L}}\big)=t$.

Using the same considerations as in Sec.~\ref{sec:comparing}, the fidelity bound for the $N$-photon sector, given by Eqs.~(\ref{eq:fidelityboundN}) and~(\ref{eq:gaussintN}), can be related to a bound on total variation distance, and the fidelity bound can be related to an entanglement fidelity.

\section{Conditional heterodyne statistics, the CC state~(\ref{eq:rhoCC}), and RBS-only characterization}
\label{sec:hetstats}

\subsection{Conditional heterodyne statistics}

Starting with a joint probability distribution $P(\bm{\alpha},\bm{n}_B|\bm{L})$,  we want to condition the heterodyne outcomes on photocount records at Bob's end.  The only conditioning we use below is on a fixed set of modes having no counts, but for generality here, we introduce a set $\Delta$ of photocount records and the projector onto the subset spanned by these records,
\begin{align}
\Pi_\Delta=\sum_{\bm{n}\in\Delta}|\bm{n}\rangle\langle\bm{n}|\,.
\end{align}
The set has indicator function
\begin{align}
\Delta(\bm{n})=\langle\bm{n}|\Pi_\Delta|\bm{n}\rangle=
\begin{cases}
1\,,&\bm{n}\in\bm{\Delta}\\
0\,,&\bm{n}\notin\bm{\Delta}
\end{cases}
\,.
\end{align}
We then have joint and conditional distributions for the set $\Delta$,
\begin{align}
P(\bm{\alpha},\Delta|\bm{L})&=\sum_{\bm{n}_B\in\Delta}P(\bm{\alpha},\bm{n}_B|\bm{L})=P(\bm{\alpha}|\Delta,\bm{L})P(\Delta|\bm{L})\,,\\
P(\Delta|\bm{L})&=\int d^{2M}\!\bm{\alpha}\,P(\bm{\alpha},\Delta|\bm{L})=\sum_{\bm{n}_B\in\Delta}P(\bm{n}_B|\bm{L})\,.
\end{align}

The mean value of a function $\Falpha$ of heterodyne outcomes is
\begin{align}
\big\langle\Falpha\big\rangle_\Delta
=\int d^{2M}\!\bm{\alpha}\,\Falpha P(\bm{\alpha}|\Delta,\bm{L})
=\sum_{\bm{n}_B\in\Delta}\frac{P(\bm{n}_B|\bm{L})}{P(\Delta|\bm{L})}\big\langle\Falpha\big\rangle_{\bm{n}_B}
\,,
\end{align}
where
\begin{align}
\big\langle\Falpha\big\rangle_{\bm{n}_B}=\int d^{2M}\!\bm{\alpha}\,\Falpha P(\bm{\alpha}|\bm{n}_B,\bm{L})\,.
\end{align}
A good way to write this is
\begin{align}
P(\Delta|\bm{L})\big\langle\Falpha\big\rangle_\Delta=\sum_{\bm{n}_B\in\Delta}P(\bm{n}_B|\bm{L})\big\langle\Falpha\big\rangle_{\bm{n}_B}\,.
\end{align}

We are generally interested in cases where $P(\Delta|\bm{L})\simeq 1-c\chi^2\simeq1-c/\sqrt M$, for some constant $c$, meaning that the probability $P(\Delta|\bm{L})$ of the record set $\Delta$ goes to 1 as $M$ goes to $\infty$, thus allowing us to use essentially all trials in the characterization.  Record sets that have this property are those that have no counts for the modes in a string ${\tt i}=i_1\ldots i_r$ and include all counts for the other modes; we denote this set by $\Delta=0_{{\tt i}}$.  Notice that changing the conditioning for just one mode in the string ${\tt i}$ from no count to a single count changes the scaling of $P(\Delta|\bm{L})$ to $\chi^2\simeq1/\sqrt M$, thus making this kind of conditioning essentially useless for characterization runs.

The probabilities that control our characterization runs are those given in Eqs.~(\ref{eq:jointPCL}), (\ref{eq:PalphanBL}), and (\ref{eq:bigPoisson2}).  We can put these together in a little bit different way,
\begin{align}\label{eq:jointPCL2}
P_{\textrm{C}}(\bm{\alpha},\bm{n}_B|\bm{L})
=P_{\textrm{C}}(\bm{\alpha}|\bm{n}_B,\bm{L})P(\bm{n}_B|\bm{L})
=\frac{(1-\chi^2)^M}{\pi^M}
e^{-\bm{\alpha}^*\ourmatrix\bm{\alpha}^T}\prod_{i=1}^{M}\frac{\big\vert\chi^2\bm{\alpha}^*\bm{L}_i\bm{L}_i^\dagger\bm{\alpha}^T\big\vert^{n_{B,i}}}{n_{B,i}!}
\,,
\end{align}
where
\begin{align}
\ourmatrix
=(1-\chi^2)\bm{I}+\chi^2\sum_{i=1}^M\bm{L}_i\bm{L}_i^\dagger
\end{align}
is a positive (Hermitian) matrix and $\bm{L}_i$ is the column vector defined by Eq.~(\ref{eq:Li}).  An important, but easily verified property,
\begin{align}\label{eq:PCstar}
P_{\textrm{C}}(\bm{\alpha}^*,\bm{n}_B|\bm{L})
=P_{\textrm{C}}(\bm{\alpha},\bm{n}_B|\bm{L}^*)\,,
\end{align}
says that conjugating the heterodyne outcomes is the same as conjugating the \LON, i.e., reversing the phase of all the matrix elements in $\bm{L}$. Equation~(\ref{eq:PCstar}) implies that $P(\bm{n}_B|\bm{L})=P(\bm{n}_B|\bm{L}^*)$, so the conjugation property extends to the conditional probabilities,
\begin{align}
P_{\textrm{C}}(\bm{\alpha}^*|\bm{n}_B,\bm{L})=P_{\textrm{C}}(\bm{\alpha}|\bm{n}_B,\bm{L}^*)\,.
\end{align}
Moreover, these properties extend to any photocount record $\Delta$,
\begin{align}
P_{\textrm{C}}(\bm{\alpha}^*,\Delta|\bm{L})&=P_{\textrm{C}}(\bm{\alpha},\Delta|\bm{L}^*)\,,\\
P_{\textrm{C}}(\bm{\alpha}^*|\Delta,\bm{L})&=P_{\textrm{C}}(\bm{\alpha}|\Delta,\bm{L}^*)\,.
\end{align}

We now focus on the photocount sets $0_{{\tt i}}$ we are most interested in.  The relevant probability distributions are
\begin{align}
P_{\textrm{C}}(\bm{\alpha},0_{\tt i}|\bm{L})
&=\frac{(1-\chi^2)^M}{\pi^M}\;e^{-(1-\chi^2)|\bm{\alpha}|^2}\prod_{s=1}^r\big\vert\braket{0|\chi\bm{\alpha}^*\bm{L}_{i_s}}\big\vert^2
=\frac{(1-\chi^2)^M}{\pi^M}\;e^{-\bm{\alpha}^*\ourmatrix_{\tt i}\bm{\alpha}^T}\,,\\
P(0_{\tt i}|\bm{L})
&=\int d^{2M}\!\bm{\alpha}\,P_{\textrm{C}}(\bm{\alpha},0_{\tt i}|\bm{L})
=\frac{(1-\chi^2)^M}{\det\bm{S}_{\tt i}}\,,\\
\label{eq:PCalphaDeltai}
P_{\textrm{C}}(\bm{\alpha}|0_{\tt i},\bm{L})
&=\frac{P_{\textrm{C}}(\bm{\alpha},0_{\tt i}|\bm{L})}{P(0_{\tt i}|\bm{L})}
=\frac{\det\ourmatrix_{\tt i}}{\pi^M}
\;e^{-\bm{\alpha}^*\ourmatrix_{\tt i}\bm{\alpha}^T}\,,
\end{align}
where
\begin{align}
\ourmatrix_{\tt i}
=(1-\chi^2)\bm{I}+\chi^2\sum_{s=1}^r\bm{L}_{i_s}\bm{L}_{i_s}^\dagger
\end{align}
is a positive (Hermitian) matrix.  These distributions and the matrix $\bm{S}_{\tt i}$ are generalizations of the distributions found in Eqs.~(\ref{eq:PCalphai}), (\ref{eq:P0i}), and (\ref{eq:PCalpha0L}), which apply to conditioning on a single no-count mode.   For a single no-count mode, it is easy to find the inverse of the matrix $\bm{S}_i$ of Eq.~(\ref{eq:ourMi}).  For two or a few modes, it would not be hard to to find the exact inverse of $\bm{S}_{\tt i}$, but for present purposes, it is simpler and more instructive to approximate the inverse for weak squeezing as a series,
\begin{align}\label{eq:Sttiinverse}
\begin{split}
\ourmatrix_{\tt i}^{-1}
&=\frac{1}{1-\chi^2}\left(\bm{I}+\frac{\chi^2}{1-\chi^2}\sum_{s=1}^r\bm{L}_{i_s}\bm{L}_{i_s}^\dagger\right)^{-1}\\
&=\frac{1}{1-\chi^2}\bm{I}
-\frac{\chi^2}{(1-\chi^2)^2}\sum_{s=1}^r\bm{L}_{i_s}\bm{L}_{i_s}^\dagger
+\frac{\chi^4}{(1-\chi^2)^3}\sum_{s,t=1}^r\bm{L}_{i_s}\bm{L}_{i_s}^\dagger\bm{L}_{i_t}\bm{L}_{i_t}^\dagger+\cdots\,.
\end{split}
\end{align}

\subsection{Conditional heterodyne moments}

Our characterization procedure runs on conditional heterodyne moments.  The moment-generating characteristic function,
\begin{align}
\Phi(\bm{\xi},\bm{\xi}^\dagger|\Delta,\bm{L})
=\big\langle e^{\bm{\xi}\bm{\alpha}^\dagger-\bm{\alpha}\bm{\xi}^\dagger}\big\rangle
=\int d^{2M}\!\bm{\alpha}\,P(\bm{\alpha}|\Delta,\bm{L})e^{\bm{\xi}\bm{\alpha}^\dagger-\bm{\alpha}\bm{\xi}^\dagger}
\,,
\end{align}
 is the Fourier transform of the relevant distribution and corresponds to $\Falpha=e^{\bm{\xi}\bm{\alpha}^\dagger-\bm{\alpha}\bm{\xi}^\dagger}$.  The characteristic function generates moments via its Taylor expansion about $\bm{\xi}=\bm{\xi}^\dagger=0$,
\begin{align}\label{eq:Phimoments}
\begin{split}
\Phi(\bm{\xi},\bm{\xi}^\dagger|\Delta,\bm{L})
&=\sum_{n,m=0}^\infty
\frac{(-1)^n}{n!\,m!}
\big\langle(\bm{\alpha}\bm{\xi}^\dagger)^n(\bm{\xi}\bm{\alpha}^\dagger)^m\big\rangle_\Delta\\
&=\sum_{n,m}\frac{(-1)^n}{n!\,m!}
\bigg\langle\sum_{n_1,\ldots,n_m}\frac{n!}{n_1!\cdots n_M!}(\alpha_1\xi_1^*)^{n_1}\cdots(\alpha_M\xi_M^*)^{n_M}\\
&\qquad\qquad\qquad\qquad\times
\sum_{m_1,\ldots,m_M}
\frac{m!}{m_1!\cdots m_M!}(\alpha_1^*\xi_1)^{m_1}\cdots(\alpha_M^*\xi_M)^{m_M}\bigg\rangle_\Delta\\
&=\sum_{\textstyle{n_1,\ldots,n_M\atop m_1,\ldots,m_M}}
\frac{(-1)^n}{n_1!\cdots n_M!\,m_1!\cdots m_M!}
\Big\langle\alpha_1^{n_1}(\alpha_1^*)^{m_1}\cdots\alpha_M^{n_M}(\alpha_M^*)^{m_M}\Big\rangle_\Delta
(\xi^*_1)^{n_1}\cdots(\xi^*_M)^{n_M}\xi_1^{m_1}\cdots\xi_M^{m_M}
\,;
\end{split}
\end{align}
thus the heterodyne moments are given by
\begin{align}
\Big\langle\alpha_1^{n_1}(\alpha_1^*)^{m_1}\cdots\alpha_M^{n_M}(\alpha_M^*)^{m_M}\Big\rangle_\Delta
=(-1)^n\left.\frac{\partial^{n+m}\Phi(\bm{\xi},\bm{\xi}^\dagger|\Delta,\bm{L})}
{\partial(\xi_1^*)^{n_1}\cdots\partial(\xi_M^*)^{n_M}\partial\xi_1^{m_1}\cdots\partial\xi_M^{m_M}}\right|_{\bm{\xi}=\bm{\xi}^*=\bm0}
\,.
\end{align}
For the Gaussian distribution of Eq.~(\ref{eq:PCalphaDeltai}), the characteristic function is
\begin{align}
\Phi_{\textrm{C}}(\bm{\xi},\bm{\xi}^\dagger|0_{\tt i},\bm{L})
=\frac{\det\ourmatrix_{\tt i}}{\pi^M}
\int d^{2M}\!\bm{\alpha}\,e^{-\bm{\alpha}\ourmatrix_{\tt i}^*\bm{\alpha}^\dagger}e^{\bm{\xi}\bm{\alpha}^\dagger-\bm{\alpha}\bm{\xi}^\dagger}
=e^{-\bm{\xi}(\bm{S}_{\tt i}^*)^{-1}\bm{\xi}^\dagger}
=e^{-\bm{\xi}^*\bm{S}_{\tt i}^{-1}\bm{\xi}^T}\,.
\end{align}

For any Gaussian distribution, characterized by a positive, Hermitian matrix $\bm{A}$ in the characteristic function, the characteristic function expands as
\begin{align}
\begin{split}
\Phi(\bm{\xi},\bm{\xi}^\dagger)
&=e^{-\bm{\xi}^*\bm{A}\bm{\xi}^T}\\
&=\sum_{n=0}^\infty\frac{(-1)^n}{n!}(\bm{\xi}^*\bm{A}\bm{\xi}^T)^n\\
&=\sum_{n=0}^\infty\frac{(-1)^n}{n!}
\sum_{\textstyle{j_1,\ldots,j_n\atop k_1,\ldots,k_n}}\xi_{j_1}^*\cdots\xi_{j_n}^*\xi_{k_1}\cdots\xi_{k_n}A_{j_1k_1}\cdots A_{j_nk_n}\\
&=\sum_{\textstyle{n_1,\ldots,n_M\atop m_1,\ldots,m_M}}
\bigg(\frac{(-1)^n\delta_{nm}}{n!}\sum_{{\tt j},{\tt k}}A_{j_1k_1}\cdots A_{j_nk_n}\bigg)
(\xi_1^*)^{n_1}\cdots(\xi_M^*)^{n_M}\xi_1^{m_1}\cdots\xi_M^{m_M}\,.
\end{split}
\end{align}
where
\begin{align}
n=\sum_j n_j\,,\qquad m=\sum_j m_j\,,
\end{align}
with $n=m$ enforced by the Kronecker delta in the sum.  The sums over products of matrix elements of $\bm{A}$ are over strings ${\tt j}=j_1\ldots j_n$ containing $n_1$ 1s, $n_2$ 2s, and on up to $n_M$ $M$s, the number of such strings being $n!/n_1!\cdots n_M!$, and, similarly, over strings ${\tt k}=k_1\ldots k_n$ containing $m_1$ 1s, $m_2$ 2s, and on up to $m_M$ $M$s, the number of such strings being $m!/m_1!\cdots m_M!$.  Comparison with Eq.~(\ref{eq:Phimoments}) shows that the heterodyne moments are given by
\begin{align}
\Big\langle\alpha_1^{n_1}(\alpha_1^*)^{m_1}\cdots\alpha_M^{n_M}(\alpha_M^*)^{m_M}\Big\rangle
=\delta_{nm}\frac{n_1!\cdots n_M!\,m_1!\cdots m_M!}{n!}
\sum_{{\tt j},{\tt k}}A_{j_1k_1}\cdots A_{j_nk_n}\,.
\end{align}
We can simplify this expression by noting that for every string ${\tt j}$, the sum over ${\tt k}$ produces the same result.  This allows us to use a standard string ${\tt j}$ (e.g., the string that has $n_1$ 1s followed by $n_2$ 2s up to $n_M$ $M$s) and thus to reduce the double sum to a single sum, giving
\begin{align}
\Big\langle\alpha_1^{n_1}(\alpha_1^*)^{m_1}\cdots\alpha_M^{n_M}(\alpha_M^*)^{m_M}\Big\rangle
=\delta_{nm}m_1!\cdots m_M!
\sum_{{\tt k}}A_{j_1k_1}\cdots A_{j_nk_n}\,.
\end{align}
Furthermore, if we restrict to the nonzero moments, i.e., those that have $n=m$, we can write the moment in a different way and turn the sum into a sum over all permutations $\pi$ of $n$ elements:
\begin{align}
\big\langle\alpha_{j_1}\cdots\alpha_{j_n}\alpha_{k_1}^*\cdots\alpha_{k_n}^*\big\rangle=\sum_{\pi\in S_n}A_{j_1,\pi(k_1)}\cdots A_{j_n,\pi(k_n)}\,.
\end{align}

For our protocol using two-mode squeezed-vacuum inputs, the heterodyne moments, conditioned on the photocount record $0_{\tt i}$, are obtained by setting $A_{jk}=(S_{\tt i}^{-1})_{jk}$, giving
\begin{align}
\big\langle\alpha_{j_1}\cdots\alpha_{j_n}\alpha_{k_1}^*\cdots\alpha_{k_n}^*\big\rangle_{{\tt i},\textrm{C}}
=\sum_{\pi\in S_n}(S_{\tt i}^{-1})_{j_1,\pi(k_1)}\cdots(S_{\tt i}^{-1})_{j_n,\pi(k_n)}\,.
\end{align}
The characterization protocol developed in Sec.~\ref{sec:characterizationdetail} relies on the first-order coherence, i.e., amplitude interference, of the heterodyne second moments~(\ref{eq:alphajki}), which are conditioned on a single no-count mode, i.e., the moments $\langle\alpha_j\alpha_k^*\rangle_{i,\textrm{C}}=(S_i^{-1})_{jk}$.  These second moments are sufficient to reconstruct the entire transfer matrix $\bm{L}$, so there is no need either for conditioning on more no-count modes or for considering higher moments, which express higher-order coherence between the complex amplitudes of Alice's modes.

The ultimate reason for introducing the general formalism for heterodyne moments, aside from completeness, is to analyze characterization using the classical-classical (CC) state~(\ref{eq:rhoCC}), a task to which we now turn.  Just before doing so, notice that we now decorate the moments derived ultimately from the distribution $P_{\textrm{C}}(\bm{\alpha}|\Delta,\bm{L})$ with a subscript C, this to distinguish them from the moments derived similarly from the CC state, to which we attach a subscript CC.

\subsection{Characterization using CC state}
\label{sec:CCcharacterization}

The classical-classical (CC) state~(\ref{eq:rhoCC}) has perfect, but purely classical photon-number correlations between the $A$ modes and the $B$ modes; thus it can be used for randomized boson sampling in exactly the same way as the two-mode squeezed-vacuum input $\ket{\Psi_{AB}}$ of Eq.~(\ref{eq:PsiAB}).  For both these states, the marginal state of either the $A$ or $B$ modes is the thermal state.  Hence, the information available from the marginal photocount records at Bob's end is the same for both inputs, and this is the ability to read out the output loss matrix $\bm{L}^\dagger\bm{L}$, as discussed in Sec.~\ref{sec:Bobstats}.  What the quantum entanglement in $\ket{\Psi_{AB}}$ enables is the characterization protocol based on first-order coherence of the complex amplitudes of Alice's modes.  In this section we investigate what information is contained in the heterodyne statistics of the CC state---and where one has to look to find this information.

A quite interesting additional question is what kind of characterization can be done with Alice's conditional photostatistics in RBS runs.  For this question, it doesn't matter whether the input is two-mode squeezed vacuum or the CC~state.  It turns out, as we confirm in App.~\ref{sec:RBSonly}, that the characterization achievable using only the RBS runs is precisely what can be achieved using the conditional heterodyne statistics derived from the CC~state.  This serves as additional motivation to suffer through the analysis presented in this subsection.

What we are interested in now is the joint probability for heterodyne outcomes and a photocount record at Bob's end, which can be expressed in terms of a phase randomization of the corresponding joint probability~(\ref{eq:jointPCL2}) for $\ket{\Psi_{AB}}$,
\begin{align}
\begin{split}
P_{\text{CC}}(\bm{\alpha},\bm{n}_B|\bm{L})
&=\frac{1}{\pi^M}
\big\langle\bm{\alpha},\bm{n}_B
\big\vert\mathcal{I}_A\otimes\mathcal{E}_{\bm{L}}(\rho_{\text{CC}})\big\vert\bm{\alpha},\bm{n}_B\big\rangle\\
&=P_{\text{C}}(\bm{\alpha})
\big\langle\bm{n}_B\big\vert\mathcal{E}_{\bm{L}}(\rho_{\text{CC}|\bm{\alpha}})\big\vert\bm{n}_B\big\rangle\\
&=\int\frac{d\bm{\Phi}}{(2\pi)^M}\,P_{\text{C}}(\bm{\alpha}e^{i\bm{\Phi}})
\Big\vert\big\langle\bm{n}_B\big\vert\chi(\bm{\alpha}e^{i\bm{\Phi}})^*\bm{L}\big\rangle\Big\vert^2\\
&=\int\frac{d\bm{\Phi}}{(2\pi)^M}\,P_{\textrm{C}}\big(\bm{\alpha}e^{i\bm{\Phi}},\bm{n}_B\big|\bm{L}\big)\,.
\end{split}
\end{align}
Here we project $\rho_{\textrm{CC}}$ onto heterodyne outcome $\bm\alpha$ to obtain the input state $\rho_{\textrm{CC}|\bm{\alpha}}$ to Bob's modes [see Eq.~(\ref{eq:rhoCChet})]; we use Eq.~(\ref{eq:rhoCCalpha}) to write $\rho_{\textrm{CC}|\bm{\alpha}}$ in terms of a $P$ function, which is a uniform distribution in the modal phases; and we use that the probability $P_{\textrm{C}}(\bm{\alpha})$ for heterodyne outcome $\bm\alpha$ [see Eq.~(\ref{eq:probCalpha})] is invariant under modal phase changes.  The conditional probability for heterodyne outcomes, given Bob's photocount record, is
\begin{align}\label{eq:PCCPC}
\begin{split}
P_{\text{CC}}(\bm{\alpha}|\bm{n}_B,\bm{L})
&=\frac{P_{\text{CC}}(\bm{\alpha},\bm{n}_B|\bm{L})}{P(\bm{n}_B|\bm{L})}\\
&=\int\frac{d\bm{\Phi}}{(2\pi)^M}\,P_{\textrm{C}}
\big(\bm{\alpha}e^{i\bm{\Phi}}\big|\bm{n}_B,\bm{L}\big)\\
&=\frac{P_{\text{C}}(\bm{\alpha})}{P(\bm{n}_B|\bm{L})}
\int\frac{d\bm{\Phi}}{(2\pi)^M}\,
\Big\vert\big\langle\bm{n}_B\big\vert\chi(\bm{\alpha}e^{i\bm{\Phi}})^*\bm{L}\big\rangle\Big\vert^2\,.
\end{split}
\end{align}
These are formal ways of writing what we already know, that these probabilities are obtained from those for the two-mode squeezed-vacuum input by randomizing the phases of the heterodyne outcomes.  They imply that
\begin{align}\label{eq:PCCprops}
P_{\text{CC}}(\bm{\alpha},\bm{n}_B|\bm{L})=P_{\text{CC}}(\bm{\alpha}e^{i\bm{\Phi}},\bm{n}_B|\bm{L})\,.
\end{align}

One thing to note at this point is the freedom we have to adjust phases in the \LON.  In the case of two-mode squeezed-vacuum input, choosing the squeezing parameter $\chi$ to be real sets relative phases between Alice's modes and Bob's modes and forbids further phase changes of the modes at the input to Bob's \LON; the remaining phase freedom is the ability to absorb phases into the definition of the output modes.  In contrast, when using $\rho_{\textrm{CC}}$ as input, there is no phase relation between Alice's modes and Bob's modes, so we have the freedom to absorb phases into both the input and the output modes of the \LON.  This means that we can choose one element of each row and one element of each column of $\bm{L}$ to be real and nonnegative.

The crucial property of the CC joint probability follows from Eqs.~(\ref{eq:PCstar}), (\ref{eq:PCCPC}), and (\ref{eq:PCCprops}):
\begin{align}
\begin{split}
P_{\text{CC}}(\bm{\alpha},\bm{n}_B|\bm{L})
&=P_{\text{CC}}(\bm{\alpha}^*,\bm{n}_B|\bm{L})\\
&=\int\frac{d\bm{\Phi}}{(2\pi)^M}\,P_{\textrm{C}}\big(\bm{\alpha}^*e^{i\bm{\Phi}},\bm{n}_B\big|\bm{L}\big)\\
&=\int\frac{d\bm{\Phi}}{(2\pi)^M}\,P_{\textrm{C}}\big(\bm{\alpha}e^{-i\bm{\Phi}},\bm{n}_B\big|\bm{L}^*\big)\\
&=\int\frac{d\bm{\Phi}}{(2\pi)^M}\,P_{\textrm{C}}\big(\bm{\alpha}e^{i\bm{\Phi}},\bm{n}_B\big|\bm{L}^*\big)\\
&=P_{\text{CC}}(\bm{\alpha},\bm{n}_B|\bm{L}^*)\,.
\end{split}
\end{align}
This property is inherited by the conditional probabilities, i.e.,  $P_{\text{CC}}(\bm{\alpha}|\bm{n}_B,\bm{L})=P_{\text{CC}}(\bm{\alpha}|\bm{n}_B,\bm{L}^*)$, and hence by the heterodyne probabilities for photocount record sets $\Delta$,
\begin{align}
P_{\text{CC}}(\bm{\alpha},\Delta|\bm{L})&=P_{\text{CC}}(\bm{\alpha},\Delta|\bm{L}^*)\,,\\
P_{\text{CC}}(\bm{\alpha}|\Delta,\bm{L})&=P_{\text{CC}}(\bm{\alpha}|\Delta,\bm{L}^*)\,.
\end{align}
What this property means is that the heterodyne statistics with CC input cannot distinguish the transfer matrix from its conjugate, i.e., from changing the sign of the phase for all the matrix elements in the transfer matrix.  We remind the reader that $P_{\text{CC}}(\bm{\alpha}|\Delta,\bm{L})$ is the Husimi $Q$ distribution for the state of Alice's modes conditioned on the photocount record $\Delta$.

We stress that this conclusion is more general than for the CC state: if the input states to a \LON\ and the output measurements are phase symmetric, the output probability distribution is invariant under exchanging $\bm{L}$ for $\bm{L}^*$.  To show this, one expresses the phase-symmetric input state in terms of phase-randomized coherent states with either a regular or singular weight function, which is the phase-averaged $P$ function.

The remaining question is whether the conditional heterodyne statistics with CC input contain all the information about the transfer matrix except this global phase reversal.  To answer this question, we look at the conditional heterodyne moments.  The key point in doing so is that the mean value of any function $\Falpha$ of the heterodyne outcomes using the CC state is obtained by randomizing the phases of the heterodyne outcomes in the corresponding mean value using the two-mode squeezed-vacuum~input:
\begin{align}
\begin{split}
\big\langle\Falpha\big\rangle_{\Delta,\text{CC}}
&=\int d^{2M}\!\bm{\alpha}\,\Falpha P_{\text{CC}}(\bm{\alpha}|\Delta,\bm{L})\\
&=\int\frac{d\bm{\Phi}}{(2\pi)^M}\int d^{2M}\!\bm{\alpha}\,\Falpha P_{\textrm{C}}\big(\bm{\alpha}e^{i\bm{\Phi}}\big|\Delta,\bm{L}\big)\\
&=\int\frac{d\bm{\Phi}}{(2\pi)^M}\int d^{2M}\!\bm{\alpha}'\,
F(\bm{\alpha}'e^{-i\bm{\Phi}},e^{i\bm{\Phi}}\bm{\alpha}^{\prime\dagger}) P_{\textrm{C}}\big(\bm{\alpha}'\big|\Delta,\bm{L}\big)\\
&=\int d^{2M}\!\bm{\alpha}\,P_{\textrm{C}}\big(\bm{\alpha}\big|\Delta,\bm{L}\big)
\int\frac{d\bm{\Phi}}{(2\pi)^M}\,F(\bm{\alpha}e^{-i\bm{\Phi}},e^{i\bm{\Phi}}\bm{\alpha}^\dagger)\\
&=\big\langle\Galpha\big\rangle_{\Delta,\text{C}}
\,.
\end{split}
\end{align}
Here
\begin{align}
\Galpha=\int\frac{d\bm{\Phi}}{(2\pi)^M}\,F(\bm{\alpha}e^{-i\bm{\Phi}},e^{i\bm{\Phi}}\bm{\alpha}^\dagger)\,.
\end{align}

For moments, which have $\Falpha=\alpha_1^{n_1}(\alpha_1^*)^{m_1}\cdots\alpha_M^{n_M}(\alpha_M^*)^{m_M}$,
\begin{align}
\begin{split}
\Galpha
&=\int\frac{d\bm{\Phi}}{(2\pi)^M}\,\alpha_1^{n_1}(\alpha_1^*)^{m_1}e^{-i(n_1-m_1)\phi_1}\cdots\alpha_M^{n_M}(\alpha_M^*)^{m_M}e^{-i(n_M-m_M)\phi_M}\\
&=\delta_{n_1m_1}\cdots\delta_{n_Mm_M}|\alpha_1|^{2n_1}\cdots|\alpha_M|^{2n_M}\,,
\end{split}
\end{align}
which leads to
\begin{align}\label{eq:condhetDeltaCC}
\big\langle\alpha_1^{n_1}(\alpha_1^*)^{m_1}\cdots\alpha_M^{n_M}(\alpha_M^*)^{m_M}\big\rangle_{\Delta,\text{CC}}
=\delta_{n_1,m_1}\cdots\delta_{n_M,m_M}\big\langle|\alpha_1|^{2n_1}\cdots|\alpha_M|^{2n_M}\big\rangle_{\Delta,\text{C}}\,.
\end{align}
It is clear what is lost by going to the classical-classical input state: one can only get at the ``diagonal'' moments of those available from the squeezed-vacuum input; these diagonal moments express intensity correlations, but lack entirely the amplitude interference that is expressed in the ``off-diagonal'' moments.

When conditioning on a single no-count mode~$i$, i.e., for the record set $\Delta=0_i$, it is obvious that the first-order coherence on
which our characterization protocol operates is missing when one uses the CC state:
\begin{align}
\langle\alpha_j\alpha^*_k\rangle_{i,\text{CC}}=\delta_{jk}\langle|\alpha_j|^2\rangle_{i,\text{C}}
=\delta_{jk}(S_i^{-1})_{jj}
=\frac{\delta_{jk}}{1-\chi^2}\bigg(1-\frac{\chi^2|L_{ji}|^2}{1-\chi^2(1-\ell_i^{\,2})}\bigg)\,.
\end{align}
What can be determined from these second moments are the magnitudes of the matrix elements of the transfer matrix, with nothing about the phases of the matrix elements.  What is not so obvious is that this is not a special property of the second moments, but for a single no-count mode, a result that holds for all moments.  The easiest way to see this is to work directly with probability distributions,
\begin{align}
\begin{split}
P_{\text{CC}}(\bm{\alpha}|0_i,\bm{L})
&=\int\frac{d\bm{\Phi}}{(2\pi)^M}\,P_{\textrm{C}}\big(\bm{\alpha}e^{i\bm{\Phi}}\big|0_i,\bm{L}\big)\\
&=\frac{\det\ourmatrix_i}{\pi^M}\int\frac{d\bm{\Phi}}{(2\pi)^M}\,e^{-\bm{\alpha}^*e^{-i\bm{\Phi}}\ourmatrix_i e^{i\bm{\Phi}}\bm{\alpha}^T}\\
&=\frac{\det\ourmatrix_i}{\pi^M}e^{-(1-\chi^2)\bm{\alpha}^*\bm{\alpha}^T}
\int\frac{d\bm{\Phi}}{(2\pi)^M}\,
\exp\!\big({-}\chi^2\bm{\alpha}^*e^{-i\bm{\Phi}}\bm{L}_i\bm{L}_i^\dagger e^{i\bm{\Phi}}\bm{\alpha}^T\big)\,.
\end{split}
\end{align}
Letting
\begin{align}
L_{ji}=|L_{ji}|e^{i\theta_{ji}}\,,
\end{align}
we can transform the phase integral,
\begin{align}
\begin{split}
&\int\frac{d\bm{\Phi}}{(2\pi)^M}\,
\exp\!\big({-}\chi^2\bm{\alpha}^*e^{-i\bm{\Phi}}\bm{L}_i\bm{L}_i^\dagger e^{i\bm{\Phi}}\bm{\alpha}^T\big)\\
&\qquad=\int\frac{d\phi_1\cdots d\phi_M}{(2\pi)^M}\,
\exp\!\Bigg({-}\chi^2\bigg(\sum_j\alpha_j^*|L_{ji}|e^{-i(\phi_j+\theta_{ji})}\bigg)\bigg(\sum_k\alpha_k|L_{ki}|e^{i(\phi_k+\theta_{ki})}\bigg)\Bigg)\\
&\qquad=\int\frac{d\phi'_1\cdots d\phi'_M}{(2\pi)^M}\,
\exp\!\Bigg({-}\chi^2\bigg(\sum_j\alpha_j^*|L_{ji}|e^{-i\phi'_j}\bigg)\bigg(\sum_k\alpha_k|L_{ki}|e^{i\phi'_k}\bigg)\Bigg)\\
&\qquad=\int\frac{d\phi_1\cdots d\phi_M}{(2\pi)^M}\,
\exp\!\bigg({-}\chi^2\sum_{j,k}\alpha_j^*e^{-i\phi_j}|L_{ji}||L_{ki}|e^{i\phi_k}\alpha_k\bigg)\,,
\end{split}
\end{align}
where $\phi'_j=\phi_j+\theta_{ji}$.  The integral over phases certainly restricts the moments to the ``diagonal'' moments, but this form makes clear that the CC probability~$P_{\text{CC}}(\bm{\alpha}|0_i,\bm{L})$, conditioned on a single no-count mode, depends only on the magnitudes, not the phases of the matrix elements of the transfer matrix.

To get phase information requires conditioning on more than one no-count mode.  Consider the second moments for the record set $\Delta=0_{\tt i}$,
\begin{align}\label{eq:jkCC}
\begin{split}
\langle\alpha_j\alpha^*_k\rangle_{{\tt i},\text{CC}}
&=\delta_{jk}\langle|\alpha_j|^2\rangle_{{\tt i},\text{C}}\\
&=\delta_{jk}(S_{\tt i}^{-1})_{jj}\\
&=\frac{1}{1-\chi^2}
-\frac{\chi^2}{(1-\chi^2)^2}\sum_{s=1}^r|L_{ji_s}|^2
+\frac{\chi^4}{(1-\chi^2)^3}\sum_{l=1}^M\sum_{s,t=1}^rL_{ji_s}L_{li_s}^*L_{li_t}L_{ji_t}^*+\cdots\,,
\end{split}
\end{align}
where we use Eq.~(\ref{eq:Sttiinverse}) for $\bm{S}_{\tt i}^{-1}$.  At order $\chi^2$, there is no phase information, but there is phase information at order $\chi^4$ and higher.  The fourth moments get at this same phase information better,
\begin{align}
\begin{split}
\langle\alpha_j\alpha_k\alpha_l^*\alpha_m^*\rangle_{{\tt i},\text{CC}}
&=\delta_{jk}\delta_{jl}\delta_{km}\langle|\alpha_j|^4\rangle_{{\tt i},\textrm{C}}
+(1-\delta_{jk})(\delta_{jl}\delta_{km}+\delta_{jm}\delta_{kl})
\langle|\alpha_j|^2|\alpha_k|^2\rangle_{{\tt i},\textrm{C}}\\
&=2\delta_{jk}\delta_{jl}\delta_{km}(S_{\tt i}^{-1})_{jj}^2
+(1-\delta_{jk})(\delta_{jl}\delta_{km}+\delta_{jm}\delta_{kl})
\big[(S_{\tt i}^{-1})_{jj}(S_{\tt i}^{-1})_{kk}+\big|(S_{\tt i}^{-1})_{jk}\big|^2\,\big]\,,
\end{split}
\end{align}
particularly the very last term.  All the terms provide phase information, but only at order $\chi^4$ and higher.  To see what can be done, specialize to the case of just two no-count modes, one chosen to be the first mode and the other mode~$i$, i.e., ${\tt i}=1i$.  In this situation, we have for $j\ne k$,
\begin{align}
\langle|\alpha_j|^2|\alpha_k|^2\rangle_{1i,\textrm{CC}}-\langle|\alpha_j|^2\rangle_{1i,\textrm{CC}}\langle|\alpha_k|^2\rangle_{1i,\textrm{CC}}
=\big|(S_{1i}^{-1})_{jk}\big|^2
\simeq\chi^4\big\vert L_{j1}L_{k1}^*+L_{ji}L_{ki}^*\big\vert^2\,.
\end{align}

Choosing the elements of the first column, $\bm{L}_1$, to be real and nonnegative leads to, for $j\ne k$,
\begin{align}
\begin{split}
|(S_{\tt i}^{-1})_{jk}|^2
&\simeq\chi^4
\big\vert L_{j1}L_{k1}+|L_{ji}||L_{ki}|e^{i(\theta_{ji}-\theta_{ki})}\big\vert^2\\
&=\chi^4
\big[L_{j1}^2 L_{k1}^2+|L_{ji}||L_{ki}|^2+2L_{j1}L_{k1}|L_{ji}||L_{ki}|\cos(\theta_{ji}-\theta_{ki})\big]\,.
\end{split}
\end{align}
By choosing, say, the first row of the transfer matrix to be real and nonnegative, we can use this to read out, for $k=1$, the cosines of all the phases in column~$i$ and, for $k\ne1$, the cosines of all the phase differences in column~$i$.  This determines the phases in $\bm{L}_i$ up to reversing the sign of all the phases in column~$i$.  Though this leaves an independent phase reversal for each column of the transfer matrix, by considering triples of modes, ${\tt i}=1ij$, one can reduce the ambiguity to an overall phase reversal that corresponds to the inability, shown above, to distinguish the transfer matrix $\bm{L}$ from its conjugate~$\bm{L}^*$.

This procedure, though tedious, shows that one can use the intensity correlations in the conditional heterodyne statistics of the CC states to read out the transfer matrix up to an overall phase reversal, but it involves working with moments at order $\chi^4$.  Reprising the argument leading to Eq.~(\ref{eq:T}) shows that in this case, to determine the complex matrix elements of $\bm{L}$ with uncertainty $\delta$ requires $T\sim M^2/\delta^2$ characterization runs. This quadratic increase with problem size, as opposed to the linear increase for the characterization procedure based on inputting two-mode squeezed vacuum, is the ultimate cost of using the classical-classical state~$\rho_{\textrm{CC}}$.

\subsection{RBS-only characterization}
\label{sec:RBSonly}

A reward---really quite a considerable reward---for having done the analysis of heterodyne statistics and the CC state is that we can now address and answer the question of what kind of characterization can be done with Alice's conditional photostatistics in the RBS runs of our protocol.

If we are interested only in photocounting at both Alice's end and Bob's end, as we are for the RBS runs, it doesn't matter whether the initial state of Alice and Bob's modes is the two-mode squeezed-vacuum state $\ket{\Psi_{AB}}$ or the CC state~$\rho_{\textrm{CC}}$, since both these states give the same photostatistics.  Indeed, Alice's photostatistics, conditioned on Bob's photocount record $\Delta$, are derived from the conditional state of Alice's modes whose Husimi $Q$ distribution is $P_{\text{CC}}(\bm{\alpha}|\Delta,\bm{L})$.  The corresponding moments of Alice's photocounts can be written as
\begin{align}\label{eq:photomoments}
\big\langle(a_1^\dagger a_1)^{k_1}\cdots(a_M^\dagger a_M)^{k_M}\big\rangle_{\Delta,\text{CC}}\,,
\end{align}
where the subscripts indicate that the expectation value is taken in the state whose Husimi $Q$ distribution is $P_{\text{CC}}(\bm{\alpha}|\Delta,\bm{L})$.  Having the photocount moments~(\ref{eq:photomoments}) is equivalent to having the rising-factorial moments
\begin{align}
\big\langle(a_1 a_1^\dagger)_{k_1}\cdots(a_M a_M^\dagger)_{k_M}\big\rangle_{\Delta,\text{CC}}
=\big\langle a_1^{k_1}(a_1^\dagger)^{k_1}\cdots a_M^{k_M}(a_M^\dagger)^{k_M}\big\rangle_{\Delta,\text{CC}}
\end{align}
where the rising factorial is given by Pochhammer's symbol,
\begin{align}
(aa^\dagger)_k=aa^\dagger(aa^\dagger+1)\cdots(aa^\dagger+k-1)=(a^\dagger a+1)(a^\dagger a+2)\cdots(a^\dagger a+k)=a^k(a^\dagger)^k\,,
\end{align}
and where, as indicated, the rising-factorial moments are the same as antinormally-ordered photocount moments.  These antinormally ordered photocount moments are precisely the conditional heterodyne moments~(\ref{eq:condhetDeltaCC}) that are available from the CC-state:
\begin{align}
\big\langle a_1^{k_1}(a_1^\dagger)^{k_1}\cdots a_M^{k_M}(a_M^\dagger)^{k_M}\big\rangle_{\Delta,\text{CC}}
=\big\langle(\alpha_1\alpha_1^*)^{k_1}\cdots(\alpha_M\alpha_M^*)^{k_M}\big\rangle_{\Delta,\text{CC}}
=\big\langle(\alpha_1\alpha_1^*)^{k_1}\cdots(\alpha_M\alpha_M^*)^{k_M}\big\rangle_{\Delta,\text{C}}\,.
\end{align}

What this burst of equations means is that the photostatistics of the RBS runs provide exactly the same information about the transfer matrix---nothing more and nothing less---as do the conditional heterodyne statistics when the input is the CC state.  Thus the RBS-only photostatistics can be used, in principle, to reconstruct the entire transfer matrix, except for the inability to distinguish $\bm{L}$ from $\bm{L}^*$.  The reason for this is that both the RBS photostatistics and the CC conditional heterodyne statistics provide access to all of the intensity correlations, but to none of the first-order, amplitude coherence that is used by our protocol for reconstructing $\bm{L}$.  This means that the RBS photostatistics must consult higher-order moments than does our protocol and thus suffer from the reduction in efficiency of reconstruction that is discussed at the end of App.~\ref{sec:CCcharacterization}.

These conclusions deserve more attention than they are likely to receive at the end of an appendix of a long paper, so we intend to redo the analysis and present the conclusions in a more straightforward way in a separate publication, thereby giving it the prominence it deserves.

\end{document}